\documentclass[iop,apj,twocolumn]{emulateapj}
\usepackage{ctable}
\usepackage{float}
\usepackage{graphicx}
\usepackage{color}
\usepackage{amsmath}
\usepackage{ctable}
\newcommand{\kms}{\ensuremath{\mathrm{km\,s^{-1}}}}
\newcommand{\sigmastar}{\sigma_{\star}}
\newcommand{\Mcutoff}{$M_{\mathrm{low} }$} 

\newcommand{\MoLs}{\Upsilon_{\star}}

\shorttitle{Non universality of the low-mass end of the IMF}
\shortauthors{Spiniello et al.}

\begin{document}

\title{The non-universality of the low-mass end of the 
IMF is robust against the choice of SSP model}

\author{C. Spiniello\altaffilmark{1},  S.~C. Trager\altaffilmark{2},
  L.~V.~E. Koopmans\altaffilmark{2}}

\altaffiltext{1}{Max-Planck Institute for Astrophysics, Karl-Schwarzschild-Strasse 1, 8l740 Garching, Germany}
 \altaffiltext{2}{Kapteyn Astronomical Institute, University of Groningen, PO Box  800, 9700 AV Groningen, the Netherlands}

\begin{abstract}
We perform a direct comparison of  two state-of-the art single stellar 
population (SSP) models that have been used to demonstrate the non-universality 
of  the low-mass end of the Initial Mass Function (IMF) slope. 
The two public versions of the SSP models are restricted to either solar abundance patterns 
or solar metallicity,  too restrictive if one aims to disentangle elemental 
enhancements, metallicity changes and IMF variations in massive early-type galaxies  (ETGs) 
with star formation histories different from the solar neighborhood.
We define response functions (to metallicity and $\alpha$-abundance) 
to extend the parameter space of each set of models. 
We compare these extended models with a sample of  Sloan Digital 
Sky Survey (SDSS) ETGs spectra with varying velocity dispersions. 
We measure equivalent widths of optical IMF-sensitive stellar features 
 to examine the effect of the underlying model assumptions and ingredients, such as stellar libraries 
or isochrones, on the inference of the IMF slope down to  $\sim0.1\,M_{\odot}$.   
 We demonstrate that the steepening of the low-mass end of the Initial Mass Function (IMF) based 
on a non-degenerate set of spectroscopic optical  indicators is robust against the 
choice of the stellar population model. 
%and show that this leads to 
%a remarkable agreement {\bf on the IMF--$\sigma$ relation} between the two extended SSP 
%models for most optical indices.  
Although the models agree in a relative sense (i.e. both imply more bottom-heavy 
IMFs for more massive systems), we find non-negligible differences on the absolute values of 
the IMF slope inferred at each velocity dispersion by using the two different models. 
In particular, we find large inconsistency in the quantitative predictions of IMF slope
variations and abundance  patterns when sodium lines are used. 
We investigate the possible reasons for these inconsistencies. 
%However,  the strong disagreement on the Na indices between the two SSP models remains unexplained, 
%hence its use in constraining the IMF slope should be more carefully examined and be considered with caution. 
%Using one set of model that allows us to change individual elemental abundances, 
%we investigate the variation of the sodium indices with IMF 
%slope and [Na/Fe] abundance, showing that the NaD feature is very 
%sensitive to [Na/Fe] variations whereas the NaI index depends mainly on the 
%IMF slope and only weakly on elemental abundance. 
%This model suggest a clear correlation between 
%[Na/Fe] and $\sigma_{\star}$, where more massive galaxies are  Na-enhanced.

\end{abstract}


%\citet[hereafter CvD12]{Conroy2012}, and \citet[hereafter MIUSCAT]{Vazdekis2012}.
%Unfortunately a complete and  flexible set of models that allows us to investigate at the same time  
%variations in age, IMF, metallicity, [$\alpha$/Fe] and abundance patterns 
%is not currently available. 

%We show that the use of the optical indices defined in \citet{Spiniello2014}, mainly from TiO and CaH 
%molecular absorption lines, permits us to eliminate the uncertainty caused 
%by the different stellar libraries used in the synthesis process. 

\section{Introduction}
A proven technique to quantitatively study the luminous stellar content of unresolved stellar populations 
is evolutionary population synthesis, first implemented by 
Tinsley (\citealt{Tinsley1968, Tinsley1972, Tinsley1976}). 
Using stellar evolution theory and stellar spectra it is possible to derive relevant stellar-population 
parameters such as age, metallicity, the shape of the stellar Initial Mass Function (IMF), element abundance, 
and the physical state and quantity of dust of galaxies (e.g.\ \citealt{Conroy2013}).

Many of the fundamental  properties of unresolved stellar populations are encoded in their spectral energy 
distributions (SEDs), and a significant effort has been invested over the last thirty years to construct 
detailed stellar population synthesis models to extract information from the SEDs of galaxies (e.g.\ \citealt{Buzzoni1989, 
Bressan1994, Worthey1994, Bruzual1993, Bruzual2003, Leitherer1999, Maraston2005}). 
All stellar population synthesis models are based primarily on three ingredients, which determine the quality of the predictions: 
i) a prescription for the IMF, 
ii) a set of stellar evolutionary prescriptions and 
iii) one or more stellar spectral libraries, either theoretical or empirical. 
%The quality of the resulting SEDs model relies to 
%a great extent on the atmospheric parameters coverage of the library 
%(temperature, gravity and metallicity). 

Historically the IMF has often been considered universal and the same as that of the Milky Way
(e.g.\ \citealt{Kroupa2001, Bell2001,Chabrier2003,Bastian2010}). However, over the last years increasingly 
more observational and theoretical evidence based on different and independent methods 
supports the idea of a non-universal low-mass end of the IMF. 
Gravitational lensing combined with dynamics modeling (e.g., \citealt{Treu2010, Auger2010b})  
and stellar population modeling (\citealt{Spiniello2012}, 2014, hereafter S12 and S14 respectively), 
spectroscopic stellar population analysis alone (e.g., \citealt{Conroy2012b, LaBarbera2013}), and spatially resolved 
kinematics and dynamics (\citealt{Cappellari2012, Tortora2013,Lasker2013}) have all shown that the slope of 
the low-mass end of the IMF steepens with galaxy mass. 

Recently, two new stellar population models (SSP) have been developed with the specific purpose 
of studying metal-rich, old stellar populations: \citet[hereafter CvD12]{Conroy2012}, and 
\citet[hereafter MIUSCAT]{Vazdekis2012}. The aim of this paper is to 
compare the models to understand whether the recent suggestions that the IMF steepens with galaxy 
velocity dispersion (S12, S14, \citealt{Conroy2012b, Ferreras2013, LaBarbera2013}) 
merely arises from a misunderstanding of their main ingredients. 
When using these SSP models to infer the stellar populations from unresolved spectra of old, evolved galaxies, 
it is essential to demonstrate that the conclusions about and predictions 
for the galaxy parameters do not depend on the assumptions of the model itself.  
As we demonstrate below, the evidence for a steepening of the low-mass IMF based on line indices is indeed robust, 
but care needs to be taken in the use of some line indices and a wide enough parameter space must be explored to break degeneracies
between metallicity, age, elemental abundance and IMF variations.  

The paper is organized as follows: in Section~\ref{sec:ssp_overview} 
we give a brief introduction to the two sets of models and highlight their similarities and differences. 
In Section ~\ref{sec:ssp_data}  we compare each model independently with SDSS galaxies. 
In Section ~\ref{sec:imfsigma} we derive the IMF--$\sigma^{\star}$ relation using a set of IMF-sensitive features
for which the models give similar prediction.
In Section~\ref{sec:ssp_conclusion} we discuss our findings and present our conclusions. Finally,
in an Appendix we perform a direct comparison of the behavior of single absorption-line indices 
for both models, and we focus on the impact of the different model ingredients and assumptions on 
the inference of the low-mass end of the IMF slope. 

\section{Simple Stellar Population Models}
\label{sec:ssp_overview}
In this Section we provide a brief introduction to the two sets of SSP models.  
We describe the main ingredients and the space of stellar parameters 
that are explored, before proceeding in Section~\ref{sec:ssp_data} to compare each model with massive early-type galaxies (ETGs).

A more detailed comparison between the different underlying assumptions and ingredients of the two SSP models 
is presented in the Appendix. There we analyze the effect of the different isochrones, stellar libraries and different approaches in dealing with 
metallicity on the inference about the low-mass end of the IMF slope.
%In particular, we show that the use of the optical indices defined in \citet{Spiniello2014}, mainly from TiO and CaH 
%molecular absorption lines, permit us to reduce the uncertainty caused  by the different stellar libraries used in the synthesis process. 

\subsection{Conroy \& van Dokkum SSP models}
CvD12 presented new SSP models with variable abundance patterns and 
stellar IMFs, suitable to study spectra of galaxies with ages $\geq 3$ Gyr\footnote{A more recent version of the SSP 
 presented in \citet{Choi2014} is capable of fitting populations as young as 1 Gyr.}. 
The models explore variation in $[\alpha$/Fe] but all of them have solar total metallicity, 
even when synthesizing models with different abundance patterns. 
They use the combination of three different isochrones to explore the separate 
phases of stellar evolution:  
i) the Dartmouth isochrones (\citealt{Dotter2008}) for the main sequence and the red giant branch (RGB); 
ii) the Padova isochrones (\citealt{Marigo2008}) to describe AGB evolution and the horizontal branch (HB); 
iii) and the Lyon isochrones (\citealt{Chabrier1997, Baraffe1998}) for stars with masses $M\leq 0.2M_{\odot}$. 
The wavelength interval covered by the final fiducial model is $0.35\,\mu m < \lambda < 2.4\,\mu m$ 
at a resolution of $R\simeq 2000$. The CvD12 models use two separate empirical stellar 
libraries: the MILES library over the wavelength range $0.35 \,\mu m  < \lambda  < 0.74\,\mu m$ 
(\citealt{SanchezBlazquez2006}) and the IRTF library of cool stars over 
the wavelength range $0.81\, \mu m  < \lambda  <  2.4\,\mu m$ 
(\citealt{Cushing2005}), plus synthetic stellar spectra to cover the gap between these two models and to 
investigate spectral variations due to changes in individual elemental abundances. 
%including carbon, sodium, calcium, iron, and generic $\alpha$-elements.
In the version  of the code that we examined, the model allows for variations in the 
elements C, N, Na, Mg, Si, Ca, Ti, Cr, Mn, Fe, O, Ne, and S. 
We use here a more recent version of the models, presented in \citet{Conroy2013b}, that use M dwarf templates
from SDSS (\citealt{Bochanski2007}) to supplement the (very small) number of dM stars present in the MILES library.

With the isochrones and stellar libraries described above, 
CvD12 construct integrated light spectra via the equation

\begin{equation}
f (\lambda)=\int_{m_{l}}^{m_{u}(t)} \,s(\lambda, m)\,\phi(m)\,dm, 
\end{equation}
where the integral over stellar masses ranges from the hydrogen 
burning limit (assumed to be $m_{l}$ = $0.08\,M_{\odot}$) to the 
most massive star alive at time $t$. In Equation~1, $f$ is the integrated spectrum, $s$ is 
the spectrum of a single star, and $\phi(m)$ = $dN/dm$ is the IMF. 
All stars in the population are assumed to have the same metallicity and abundance pattern. 
CvD12  explore variations in age between 
$3$--$13.5$ Gyr, four different single-slope IMFs  --  a Chabrier (2003) Milky Way-like IMF with a slope of $x = 1.8$, 
hereafter referred to as the `MW IMF',  a Salpeter IMF ($x = 2.3$), and two bottom-heavy IMFs with slopes of 
$x = 3.0$ and $x = 3.5$ -- and different $\alpha$-enhancement and individual element abundances.

\subsection{The MIUSCAT SSP models}
The MIUSCAT models are an extension of stellar population synthesis 
models based on the MILES (\citealt{SanchezBlazquez2006}) and 
CaT (\citealt{Cenarro2001}) empirical stellar spectral libraries to cover the spectral 
range $0.346\,\mu m   < \lambda  <  0.947\, \mu m$. 
Moreover,  the spectral coverage is extended to the blue and to the red with the Indo-U.S. library (\citealt{Valdes2004}).
In order to determine which stars to include in the synthesis,  
they use the solar-scaled theoretical isochrones of \citet{Girardi2000}, which 
 cover a wide range of ages, and six metallicity bins. 
The Girardi isochrones include a simple synthetic prescription that incorporates the thermally-pulsing 
AGB regime (\citealt{Bertelli1994}). Moreover an improved version 
of the equation of state, new opacities from \citet{Alexander1994} and a convective overshoot scheme
 have been added to the models to improve the physics of these latest stages of stellar evolution.  
%including a simple synthetic prescription for incorporating the thermally-pulsing 
%AGB regime to the point of complete envelope ejection (\citealt{Bertelli1994}). 
%The physics of these models was updated with respect to \citet{Bertelli1994} 
%with an improved version of the equation of state, the opacities of 
%\citet{Alexander1994} and a milder convective overshoot scheme. 
The stars are then attached to the isochrones according to their 
number per mass bin, predicted from the adopted IMF. 
Different IMFs shapes are considered: the unimodal and bimodal power-law IMFs 
defined in \citet{Vazdekis1996}, and the multi-part power-law IMFs of \citet{Kroupa2001}. 
The Salpeter (1955) IMF is represented by the unimodal case with slope $\Gamma = x-1$ = $1.3$. 
Here we restrict to the unimodal power-law IMFs to perform a fair comparison with CvD12.
We use the Kroupa (2001) `universal' IMF to represent a Milky Way-like IMF for the MIUSCAT models, 
referred to hereafter as the `MW IMF'. We caution the reader that this IMF is \emph{not exactly} 
the same as the `MW IMF' used in the CvD12 models (single power-law with a slope $x=1.8$ that gives a 
$M/L$ similar to the one obtained with a standard Chabrier), 
but should be close given the similar  shapes and normalization (see, e.g., \citealt{Chabrier2005}). 
 
The lower stellar mass limit (cutoff mass) 
assumed by the MIUSCAT models, given the \citet{Girardi2000} isochrones, 
is \Mcutoff$ = 0.15\,M_{\odot}$, slightly higher than that used in CvD12.
From a spectroscopic point of view, 
the adoption of a slightly higher value of \Mcutoff \, does not have a large impact on 
the line-strength measurements, although it can have a visible contribution for some 
spectral features (see Fig.~15 in \citealt{Conroy2012}).
However, one has to keep in mind that even though stars below $\sim 0.1 M_{\odot}$ 
are almost invisible in current spectral lines, they can contribute substantially to the total stellar mass 
and number of stars for any standard IMF. 
As discussed in \citet{Barnabe2013}, the value of \Mcutoff \, is an essential parameter when 
determining $\MoLs$ from stellar population codes. 

%We show that the use of the optical indices defined in \citet{Spiniello2014}, mainly from TiO and CaH 
%molecular absorption lines, permits us to eliminate the uncertainty caused 
%by the different stellar libraries used in the synthesis process. 

\subsection{Extending the parameter space of the models}
\label{sec:responsefunction}
A significant difference between the two sets of models is the different approach to dealing with metallicity and [$\alpha$/Fe]. 
The CvD12 models use solar-metallicity isochrones even when synthesizing models with different abundance patterns. 
The total metallicity $Z$ varies from model to model, because the abundance variations of single elements are implemented at fixed [Fe/H]. 
%, which implies that the total metallicity $Z$ varies from model to model. 
The MIUSCAT models, on the other hand, do not allow the relative abundance of the $\alpha$-elements to change and 
are therefore restricted to the solar abundance pattern, although there are models with different total metallicities. 
Each MIUSCAT SSP has a fixed total metallicity, and in total MIUSCAT has six metallicity values  
($Z=0.0004, 0.001, 0.004,0.008,0.019$ and $0.03$, corresponding to $[M/H]=-1.71, -1.31, -0.71, -0.40,0.0,+0.22$). 

Therefore, the only direct comparison possible between the public versions of the models  
is at solar metallicity and solar [$\alpha $/Fe]. 
This is restrictive if one aims to disentangle elemental enhancements, metallicity changes and IMF variations. This is especially true for
very massive ETGs, which are known to be $\alpha$-enriched and to have slightly super-solar metallicity, due to star formation histories 
different from the solar neighborhood (e.g., \citealt{Peterson1976,Peletier1989,Worthey1992, Trager2000b, Arrigoni2010}).  

To resolve this, at least on a qualitative level, we extend the parameter space of each set of models in the following ways.
We take the ratio between two spectra of MIUSCAT models with same age and IMF 
slope but different total $Z$. In this way we isolate the effect of changing the total metallicity from the effect of changing 
other stellar population parameters, i.e. we construct a metallicity-response function for each given age and IMF slope. 
We then multiply this response function with the spectrum of a CvD12 model with same age and IMF 
to build a new model (SSP) that extrapolates the latter model in to a new part of parameter space (i.e covering super-solar $Z$):

\begin{equation}
\Delta Z(\tau, x)=\dfrac{\mathrm{MIU}(\tau, x, Z)}{\mathrm{MIU}(\tau, x, Z_{\odot})}
\end{equation}

\begin{equation}
\mathrm{CvD12_{ext}}(\tau,x, Z) \equiv \mathrm{CvD12}(\tau, x) \times \Delta Z (\tau, x)
\end{equation}

\noindent where $\tau$ is the age of the stellar population and $Z>Z_{\odot}$ is the super-solar metallicity value explored by the MIUSCAT models ($Z=0.03$, corresponding to  $[M/H]= +0.22$).  

In the same way, we use CvD12 models at fixed age and IMF slope to build an [$\alpha $/Fe]-response function that we then multiply
with MIUSCAT models to build models with super-solar [$\alpha $/Fe] abundances, 

\begin{equation}
\Delta [\alpha \mathrm{/Fe}] (\tau, x)=\dfrac{\mathrm{CvD12}(\tau, x, [\alpha \mathrm{/Fe}])}{\mathrm{CvD12}(age, x, [\alpha \mathrm{/Fe}]_{\odot})}
\end{equation}
\begin{equation}
\mathrm{MIU_{ext}}(\tau,x, [\alpha \mathrm{/Fe}]) \equiv \mathrm{MIU}(\tau, x) \times \Delta [\alpha \mathrm{/Fe}](\tau, x),
\end{equation}

 \noindent where the super-solar $[\alpha \mathrm{/Fe}]$ values explored by the CvD12 models are 0.2, 0.3, 0.4. 

These modified SSP models combine the flexibility of both the MIUSCAT and CvD12 models  
to predict spectra in a part of parameter space that neither of them reaches separately. 
Allowing for variation of age, metallicity and [$\alpha$/Fe] -- i.e.\, selecting IMF-dependent 
features that are age- and metallicity-independent and combining them with indices 
that depend mainly on age or mainly on element abundance --
is important to break the age-metallicity-IMF degeneracy when using 
SSP models to infer the stellar populations from unresolved galaxies spectra.

However, it is important to clarify that the model extension performed above has to be interpreted 
as a first order approximation meant to demonstrate that the IMF--$\sigma_{\star}$ relation is a robust result, 
independent of the choice of the SSP model. 
 In fact, the two SSP models differ in  many of their main ingredients and construction methods such as the isochrones, the 
interpolation scheme or the conversion of the theoretical parameters of the isochrones to observational quantities. 
Nevertheless we show in this paper that the combined models are very useful to demonstrate the
robustness of the results on the IMF variation obtained with CvD12 or MIUSCAT models separately.

\begin{figure*}  
\includegraphics[height=13.3cm]{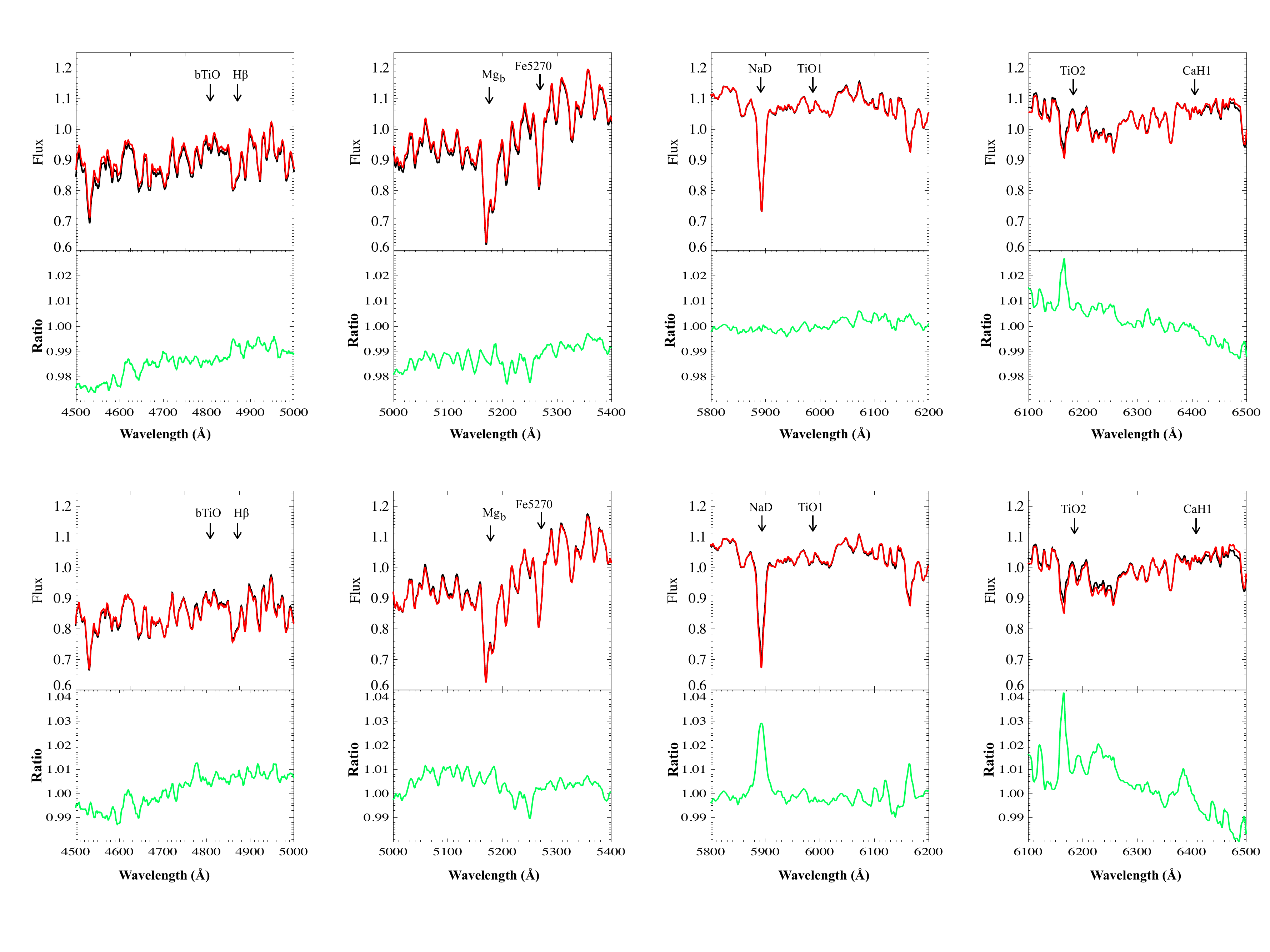}
\caption{Zoom-in of different regions of a $\mathrm{ CvD12_{ext}}$ SSP model spectrum (in black) 
and a $\mathrm{ MIU_{ext}}$ SSP model spectrum (in red) for a MW IMF (top panels) and a bottom-heavy IMF ($x=3.0$, bottom panels). 
Both models have same age  (12.5 Gyr), super-solar metallicity ($Z=0.03$) and super-solar [$\alpha$/Fe] ([$\alpha$/Fe]=$+0.2$). 
The spectra have the same resolution ($\mathrm{FHWM} =2.51$ \AA) and are plotted in unit of normalized flux. 
Some absorption lines are highlighted in the plots and the ratios between the spectra are showed in the lower green panels. 
The spectra obtained via the response-functions are very similar, especially in the case of a MW-like IMF (1\% agreement), 
demonstrating the validity of the extrapolation approach. Some differences are visible for the spectra having a bottom-heavy 
IMF. See the text for a detailed discussion. } 
\label{fig:spectra_ext}
\end{figure*}

 In Figure~\ref{fig:spectra_ext} we plot spectra of a $\mathrm{CvD12_{ext}}$ SSP model (black) 
and a $\mathrm{MIU_{ext}}$ SSP model (red) with the same age ($12.5$ Gyr),  super-solar metallicity (Z$=+0.03$), 
 super-solar [$\alpha$/Fe] ($=+0.2$)  and IMF (MW in the upper panels and $x=3.0$ in the lower panels), 
 to verify that the extrapolated versions of the CvD12 and MIUSCAT models agree for the same stellar population 
 parameters, validating our methodology.

Spectra with a MW-like IMF typically agree with $\le 1$\%, demonstrating that our approach of extrapolating one model 
using a response function obtained from the other provides mutually-consistent answers. However, some differences are visible in the 
case of a bottom-heavy IMF ($x=3.0$), especially for the NaD and TiO$2$ absorption features. 
This implies that the two sets of models give different predictions on the variations of the index strength with IMF slope, if these indices are used. 
In the following we demonstrate that this difference is caused by the different ingredients and assumptions of the two models. 
In particular we show that part, but not all, of the difference is due to the different sets 
of isochrones used in the two models. Another source of this difference is the different ways in which 
CvD12 and MIUSCAT attach stars to the isochrones (see the Appendix for more details).
Despite these differences, we show in this paper that both models predict a non-universality of the low-mass 
end of the IMF, which steepens with the stellar velocity dispersion.

\section{Comparison with data}
\label{sec:ssp_data}

We now compare the extended versions of the CvD12 and MIUSCAT models with data on ETGs and  
show that the non-universality of the low-mass end of the IMF
%--$\sigma_{\star}$ relation 
holds and appears to be robust against the choice of  SSP model in a relative sense. 
%model-independent result.
%, for which they are specifically designed.
We use a sample of SDSS spectra that has been selected  and extensively described in S12 and S14. 
The spectra were averaged (``stacked'') in five velocity-dispersion bins spread over 
$150$--$310\,\kms$ to increase the final S/N in each bin. 
%Here we also include a lower velocity-dispersion-bin with an average value of $110\,\kms$, 
%although we limit our analysis to galaxies with $\sigmastar \geq 150\,\kms$.
Below $150\,\kms$ contamination from spiral galaxies becomes larger 
and emission lines affect our results considerably. 
%To minimize spiral galaxy contamination in our ETGs sample, we look at the surface brightness models of SDSS photometric data
%and select systems for which the likelihood of a de Vaucouleurs' model is 
%larger than the likelihood of an exponential model. 
%In addition, we select systems with very low star-formation rates and 
%we set an upper limit on the redshift range ($z\leq 0.05$) to cover the wavelength range of interest. 
We refer to S14 for further details.

We perform line-strength measurements  following the approach presented 
in \citet{Trager2008}  and using their Python implementation of the algorithm, SPINDEX2. 
We measure the standard Lick indices H$\beta$, Mg$b$, Fe5270, Fe5335, NaD and  TiO$2$ 
using the definitions of Trager et al. (1998), the broadly-used [MgFe] 
combination\footnote{[MgFe] = $\sqrt{(\mathrm{Fe5270} +
\mathrm{Fe5335})/2 \times \mathrm{Mg}b}$, \citet{Gonzalez1993}}
and the IMF-sensitive set of optical indicators described in S14. 
Finally we use the modified NaI index defined in \citet{Spiniello2012} 
and the CaT index (\citealt{Cenarro2001}).
We measure  indices in the wavelength range $4000$--$8500$~\AA. We are aware that for wavelengths 
redder than $\sim 7500$~\AA, the two models make use of  different stellar libraries (see the Appendix for more details). 
All the galaxy and model spectra are convolved to a final velocity dispersion of $\sigma$ = $350\,\kms$ 
to correct for kinematic broadening before measuring line-strength. 
%We also normalize the spectra using a second order polynomial fit. 
Indices in both the galaxies and the two sets of model spectra are measured 
with the same definitions and methods.  
We do not place our indices on the zero-point system of the Lick indices. We present them 
as equivalent widths (EWs) in units of \AA, except for the molecular TiO and CaH indices, which are given in magnitudes.

\begin{figure*}  
\includegraphics[height=14cm]{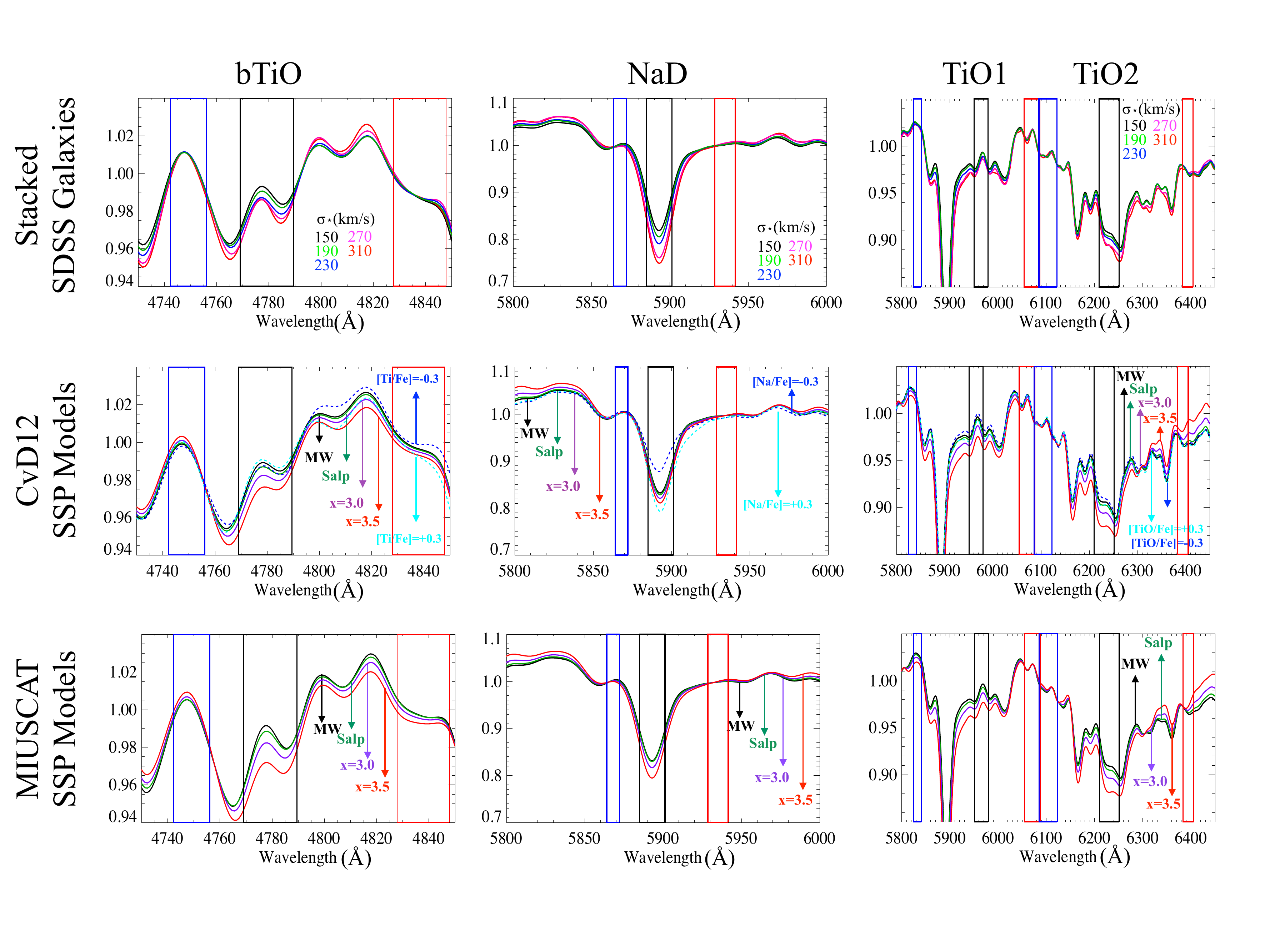}
\caption{Galaxy (top row) and SSP models (middle and bottom rows) spectra in
the regions of the bTiO, NaD, TiO$1$ and TiO$2$ absorption features. 
A clear trend of increasing EWs is visible in both data and models, although the data show a stronger variation (see text). 
The bandpasses of the indices, as well as the blue and red pseudo-continua, are shown as boxes in the plots. 
{\sl Top row:} Spectra of SDSS galaxies stacked in different velocity dispersions bins over the range $150$ -- $310\,\kms$.
{\sl Middle row:} CvD12 SSP models with age of 13.5 Gyr and solar $[\alpha$/Fe], but with different 
IMF slopes, from MW-like to very bottom-heavy slope ($x$ = $3.5$). Dotted lines show models with 
MW-like IMF and different [Ti/Fe] or [Na/Fe] abundances. 
{\sl Bottom row:} MIUSCAT SSP models with an age of 14.2 Gyr and solar metallicity, but with different IMF slopes, from the MW-like 
to an extremely bottom-heavy IMF with a slope of $x$ = $3.5$. 
All spectra are normalized to the central point of the blue and red pseudo-continua bands.}
\label{fig:ssp_spectra}
\end{figure*} 

Figure~\ref{fig:ssp_spectra} shows a zoom-in of the stacked ETGs and SSP model spectra in the bTiO, NaD,TiO$1$ and TiO$2$ regions, 
with the indices bandpass as well as the blue and red pseudo-continua bands shown as boxes. 
A clear increase of the line-strengths of all these features are visible in all panels, although in some cases (e.g. NaD) the data show 
a somewhat stronger variation, due to the fact that these indices are not only 
gravity-sensitive but also depend on individual elemental abundances and possibly age. 
For instance, for the CvD12 models, which allow a variation of the Ti and Na abundance pattern, 
we investigate whether a non-solar [Ti/Fe] abundance (left and right panels, second raw) or a non-solar [Na/Fe] abundance 
(middle panel, second raw) could significantly vary the  indices strength and explain the variation seen in the data. 
It is clear that the variation due to IMF can be mimicked by non-solar abundance. 
We therefore stress that the use of many indicators, arising from different features, is crucial to break such degeneracies (S14). 
In particular, in the case of the NaD index, the variation of the NaD EWs in the CvD12 models due to a non-solar [Na/Fe] abundance is larger 
than the variation due to IMF slope.   
The larger variation of NaD EWs due to non-solar [Na/Fe] abundance seems to be in better agreement with the variation observed for SDSS galaxies. 
We investigate this further in the Appendix, however we caution the reader that  this is a model-dependent result.

\begin{figure*}  
\includegraphics[height=4.5cm]{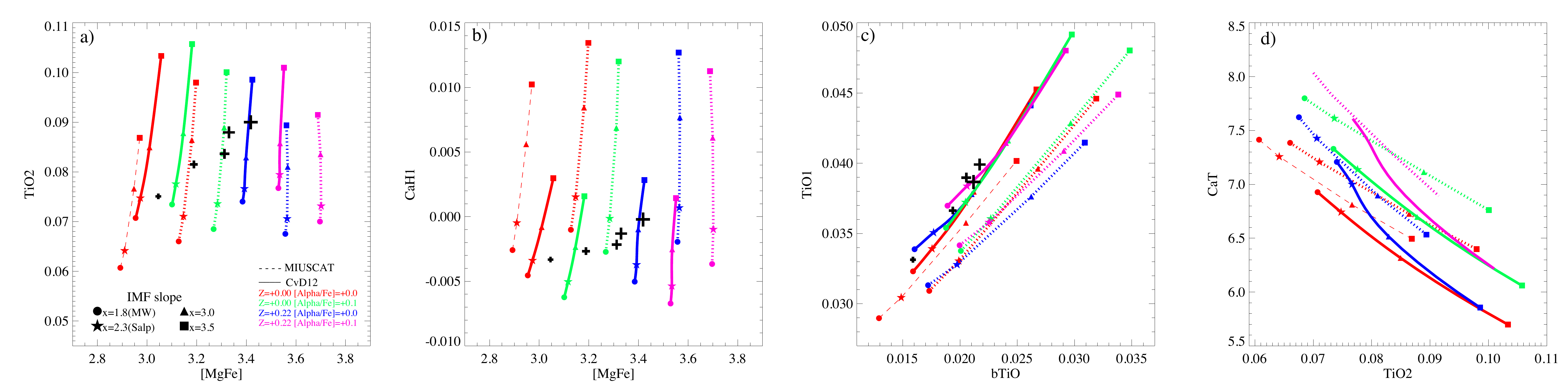}
\caption{Index-index plots of some of the most prominent IMF-sensitive absorption features in the optical regime. 
Solid lines are CvD12 SSP models and dotted lines are MIUSCAT SSP models; both models have an age of $12.5\,$Gyr.   
The dashed line is a MIUSCAT model with an age of $\sim 9.0\,$Gyr. 
Red lines are SSP models with solar metallicity and solar $[\alpha$/Fe], blue lines are SSP models with 
$Z$ = $+0.22$ and solar $[\alpha$/Fe], green lines are $\alpha$-enhanced models with solar metallicity
 and magenta lines are $\alpha$-enhanced models with super-solar total metallicity.  
%One should use super-solar and/or $\alpha$-enhanced models to compare with massive ETGs because 
%they have likely undergone a star formation histories 
%different from the solar-neighbourhood (e.g., \citealt{Peterson1976,Peletier1989,Worthey1992, Trager2000b, Arrigoni2010}).  
Symbols on each line represents different IMF slopes (see legend).
%on the same line are models with increasing IMF slope (Chabrier, Salpeter with a slope of $x$ = $2.35$, and several 
%bottom-heavy IMFswith slope between $2.8$ < $x$ < $3.8$). 
Black points with error bars are SDSS galaxies, stacked by velocity dispersions expressed in $\,\kms$. 
Bigger symbols are galaxies with larger velocity dispersions. 
On a qualitative level, these diagrams show a good agreement of the two models with the galaxies for these indices: 
a clear trend of the low-mass slope of the IMF with galaxy mass is visible in both models: the most-massive ETGs 
require an IMF slope slightly steeper than Salpeter for both SSP models. 
However, absolute values of the inferred IMF slopes for each $\sigma_{\star}$
are different (see text and Appendix for more details).} We remind the reader that TiO and CaH indices 
 are given in magnitudes.
\label{fig:ssp_data1}
\end{figure*}

In Figure~\ref{fig:ssp_data1} and Figure~\ref{fig:ssp_data2} we present index-index plots of the IMF-sensitive features for 
the two extended sets of models and the stacked SDSS galaxies. 
Remarkably, massive galaxies better match SSP models with steeper IMF slopes independently 
of the considered set of models in all panels.  Index-index plots are a useful tool to give qualitative inference 
on the stellar population parameters, but a proper statistical analysis involving a wide range of spectral features is necessary 
to break degeneracies and quantitatively constrain age, metal abundance and IMF slope, as we did in S14 via a $\chi^{2}$ minimization routine (using CvD12).
We restrict the comparison to a single SSP age ($12.5 \pm 1$ Gyr{\footnote{The two sets of models do not have a common old age: the oldest CvD12 models have ages of 11 and 13.5 Gyr respectively;  the oldest MIUSCAT models have ages of 12.5 and 14.1 Gyr.
Therefore we interpolated the CvD12 models to create 12.5 Gyr models to eliminate most of the residuals arising from age differences.}) 
with varying IMF slope, metallicity and [$\alpha$/Fe] and focus only on features with a weak age-sensitivity.  
In addition, we plot a younger MIUSCAT models ($\sim 9.0$ Gyr) with solar $\alpha$-abundance and metallicity 
to hightlight the age effect on the line-index strength variations. 
We note that in S14 we find that all galaxies (with $\sigma \geq 150\,\kms$) have old stellar populations ($>9$ Gyr). }

Very massive ETGs are known to be $\alpha$-enriched and to have slightly super-solar metallicity (see, e.g., \citealt{Worthey1994}).   
Our aim here is to compare each set of models with massive galaxy spectra bins, and we therefore plot models built using 
our response functions (see previous section)  with super-solar metallicity and super-solar [$\alpha $/Fe]. 

%in  order to compare the SSP  models with our more massive galaxy spectra bins.
We limit ourselves to a unimodal IMF slope, the only choice for the CvD12 models, but we stress that 
it may not be possible to fully constrain the detailed functional form of
the IMF (either unimodal or multi-segmented) using only index-index diagrams. 
For instance, although the mass-to-light ratios obtained in S14 via line-strength measurements 
assuming a unimodal IMF we were consistent with the results of  \citet{LaBarbera2013}, the inferred IMF slopes were not. 
These authors obtained steeper slopes when using a two-segmented
(bimodal) IMF yet found similarly good fits to the data using unimodal IMF slopes in their index-index plots. 
A note of caution should be also sounded concerning the fact that the two different SSP models
adopt a different choice for \Mcutoff, implying different results 
for the stellar mass-to-light ratios  (CvD12 use 0.08 M$_{\odot}$, while MIUSCAT  adopt 0.10M$_{\odot}$).
We clarify that here (and in previous papers) the IMF--$\sigma^{\star}$ relation has been obtained under the assumption of a 
fixed, universal lower cutoff mass (0.10 M$_{\odot}$). 

Figure~\ref{fig:ssp_data1} shows some of the IMF indicators in the optical that were shown to robustly break degeneracies in the 
SSP models between age, metallicity, abundance pattern and IMF slope in S14.  
For both models this set of indicators clearly show a steepening of the IMF slope with stellar velocity dispersion (galaxy mass), 
although zero-point shifts and differences in the absolute values of the IMF slopes are visible (see Appendix). 
The models imply a MW-like IMF for the least-massive galaxies 
($\langle \sigma_{\star}\rangle$ = $150\,\kms$), 
a Salpeter IMF for the intermediate-mass ETGs 
and possibly a bottom-heavy IMF (with $x \sim 3$) for the most massive galaxies. 
This result is fully consistent with the more-detailed analysis performed in S12, S14 and 
in completely independent studies (e.g. \citealt{LaBarbera2013}) 
but here using both the CvD12 and MIUSCAT models in a consistent way.  

Panels (a), (c) and (d) of Figure~\ref{fig:ssp_data1} show a fair agreement between the two set of models and the 
galaxies for old stellar populations, although some differences still remain.  
The [MgFe]--TiO$2$ diagram (panel a) suggests that more-massive galaxies 
require super-solar metallicity and possibly also super-solar [$\alpha $/Fe].  
The same is also visible in panel (b), but in this diagram the differences in the absolute IMF values are extreme. 
In fact, CaH1 variations with IMF slope predicted from the two sets of models are extremely different and an offset for this index is also present. 
In panel (c) and (d) the dependencies on Z and [$\alpha $/Fe] are minimal and somehow degenerate with IMF slope.  
Here the models give similar predictions for  the steepening of the IMF slope, but an offset between MIUSCAT and CvD12 models 
with the same parameters is found. 
We investigate the possible reasons for this offset in the Appendix, finding that the different isochrones used by the two models
play a non-negligible role for most of the indicators but do not solve the problem for the TiO$2$ index. 
We believe that this difference is primarily due to the different methods used in the 
CvD12 and MIUSCAT models to attach stars to the isochrones at low mass. 
%We also note that effective temperature of the isochrones ($\Delta T_{\rm eff}$) could cause some of the zero point offset but. 
 %As we demonstrated in S14,  the response of a change in $\Delta T_{\rm eff}$ is similar to the response of a change in 
%the number of dwarf stars in the galaxy.  We speculate further on the IMF slope -- $\Delta T_{\rm eff}$ 
%degeneracy in the Appendix, showing that the zero-point shift visible in most of the indicators
%could be due to differences in the temperatures of cold stars. However, to confirm this suspicion, more investigation is required. }
%(e.g. for a Salpeter IMF the TiO$2$ EW in the MIUSCAT models is $\sim 9$ percent larger than the EW in the CvD12, 
%whereas the CaH1 EW in the MIUSCAT models is $\sim 7$ percent smaller than the EW of the same index in the CvD12) 

%In particular, in the case of the TiO$2$, 
%the models do not predict the same variation of  the  EW with IMF slope, 
%even when they make use of the same set of isochrones. 

Panel (d) confirms the previously-known fact that the CaT index is almost metallicity 
independent\footnote{This is true when $[Z/{\rm H}]\geq -0.5$, 
which is the case for the giant ETGs considered here.}, as visible from the MIUSCAT models, 
while it strengthens with increasing [$\alpha$/Fe]. %The same is true for the TiO lines.  
%The CaH1 index, also almost metallicity independent, increases with steepening of the IMF slope and decreases 
%with increasing [$\alpha$/Fe], as visible in panel (c). 
%The  [$\alpha$/Fe]-IMF degeneracy can be easily broken with this particular plot, since 
%the variation of IMF slope is almost orthogonal to the [$\alpha $/Fe] enrichment.   
%Since the CvD12 models with solar [$\alpha $/Fe] provide the best match with data for all the mass bins, 
%one might argue that perhaps the IRTF library is a better choice to 
%study the low-mass star population than the CaT library of \citet{Cenarro2001} used by MIUSCAT; 
%but it would be premature to conclude this without further detailed comparison of the CaT and IRTF libraries in this region. 
The two SSP models use different libraries in the CaT wavelength range, 
but they still lead to the same prediction for the IMF variation. We therefore conclude that the 
use of different libraries is not responsible for the large disagreement visible when sodium lines are used. 

Finally, we note that the zero-point shift cannot be explained by allowing for an age offset between the two sets of models. 
A MIUSCAT model with age $\sim 9$ Gyrs (red, dashed line in Fig.\ref{fig:ssp_data1}) and with a MW-like IMF predicts a [MgFe] strength 
very similar to that predicted by CvD12 with the same IMF but $\sim 4$ Gyr older. 
However, the EWs predicted by these two models with different ages for the bTiO and the TiO2 indices 
 are in much worse agreement with respect with those predicted from models with the same age.

\subsection{The sodium features}
\begin{figure}  
\center
\includegraphics[height=11cm]{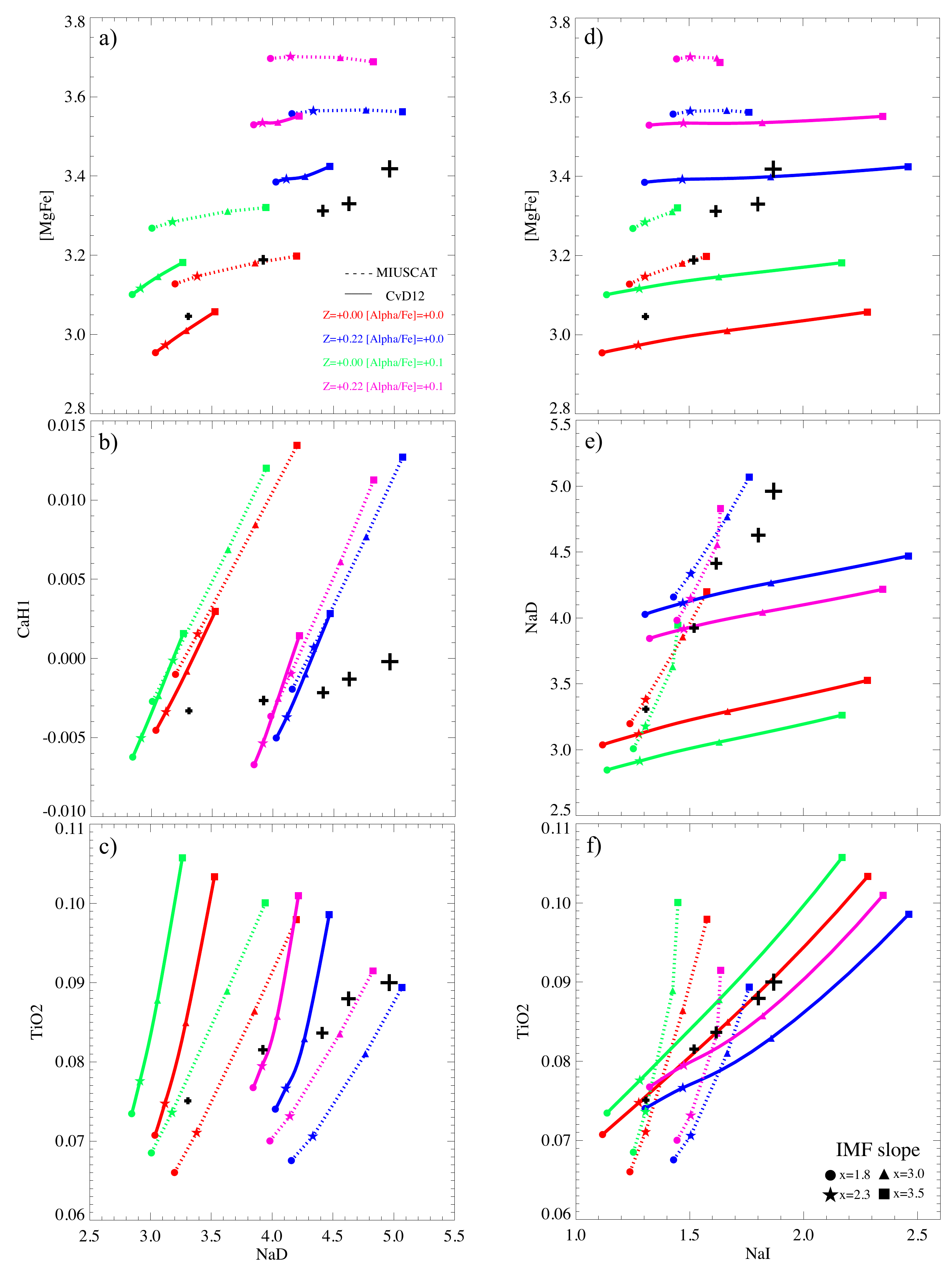}
\caption{Index-index plots of the blue sodium absorption feature NaD (left column, panels a, b and c) and the red 
sodium absorption feature NaI (right column, panels d, e and f) in the optical versus other IMF-sensitive features. 
Lines and data-points are the same as in Figure \ref{fig:ssp_data1}.
%Red lines are SSP models with solar metallicity and solar $[\alpha$/Fe], blue lines are SSP models with 
%$Z$ = $+0.22$ and solar $[\alpha$/Fe], green lines are $\alpha$-enhanced models with solar metallicity 
%and finally magenta lines are $\alpha$-enhanced models with super-solar total metallicity.  .  
%Symbols on each line represents different IMF slopes (see legend).
%Black points with error bars are SDSS galaxies, stacked by velocity dispersions expressed in $\,\kms$. 
Bigger symbols are galaxies with larger velocity dispersions. 
In these plots the two sets of models give very different predictions for IMF variations.
\textit{Panels (a), (c) and (e) :} MIUSCAT models with varying total metallicity match the data better at all $\sigmastar$ 
than CvD12 models, but the most massive bin with 
$\langle \sigma \rangle$ = $310\,\kms$ requires an extremely steep IMF with a slope of $x$ = $3.5$ (blue square), 
in disagreement with previous inferences based on lensing and stellar populations (\citealt{Spiniello2012}). 
\textit{Panel (b):} None of the models match the data at all  all $\sigmastar$ 
%unless we do not require an extremely steep $Z-\sigma$ relation
%, which would however disagree with previous results based on other indices.  
For all these panels, a sodium over-abundance with [Na/Fe] = $0.3$--$0.5$ 
(consistent with values found by \citealt{OConnell1976, Peterson1976, Carter1986, Alloin1989, Worthey1998, Worthey2011}) 
in more massive galaxies can explain the disagreement between models and the most massive galaxies bin. 
This is however a model-dependent result, obtained with CvD12.  
\textit{Panels (d) and (f):} Here MIUSCAT SSP models do not match the index strengths 
  of very massive systems, while the CvD12 models with solar metallicity and varying IMF slopes 
do without requiring an incredibly steep IMF slope for the most massive bin.  
As shown in Figure~\ref{fig:ssp_sodiumlines} and Table~\ref{tab:sodium_delta}, NaI does not strongly depend on sodium abundance.}
\label{fig:ssp_data2}
\end{figure}

In Figure~\ref{fig:ssp_data2} we plot the two sodium absorption features in the optical (the blue NaD index at $\lambda\sim5900$\AA\, 
on the left column and the redder NaI doublet at $\lambda\sim8190$\AA\, on the right) against the IMF-sensitive features previously introduced.  
In all cases, the two models have very different behavior and in similar cases they fail to match the data at all velocity dispersions. 
Specifically the CvD12 models with solar abundance pattern only match the low-mass systems in all panels with NaD 
(a, b, and c)\footnote{We confirm here the finding of S12 that for the CvD12 models the NaD indices and their trends with stellar 
mass remain unexplained, at least for the more massive systems with $\sigma \geq 200$ \kms, using models with  solar abundance pattern}. 
MIUSCAT models better match the data at all velocity dispersions, but do not predict 
the same IMF slope for the same velocity-dispersion bin in each panel. In panel (b) the IMF slope for the most massive bin 
is close to a Salpeter IMF, whereas in panel (a) and (c) MIUSCAT predicts extremely bottom-heavy IMF slopes, steeper than
 those inferred from other absorption-line indices of Figure~\ref{fig:ssp_data1}\footnote{The same result was obtained for CvD12 models in S14. 
In S14 we show that the quantitative relation between IMF slope and velocity dispersion 
inferred including the NaD index is systematically different than the relation inferred from any other combination 
of the IMF-sensitive spectral indices mentioned above.}.  
These trends also violate lensing and dynamical constraints (\citealt{Treu2010, Spiniello2011, Spiniello2012, Barnabe2013}).  

Given this situation, further investigation into the different behavior of the sodium indices in the two SSP models is necessary. 
We show in the Appendix that the NaD index is much more sensitive to [Na/Fe] abundance than to IMF slope variation in the CvD12 models. 
Moreover, if one assumes a relation between [Na/Fe] and $\sigma_{\star}$, 
in the sense that more massive galaxies are Na-enhanced,  then the CvD12 models match the data in all velocity bins. 
Although a non-solar [Na/Fe] abundance for more-massive galaxies has been reported in literature (e.g. \citealt{Jeong2013}), we 
stress that this result is model-dependent and therefore should be taken with some caution. 

%are also inconsistent with all other absorption-line index trends and further with lensing and dynamical constraints.  
%Finally, in panel (f), the CvD12 models better match the indices of galaxies at all velocity dispersions bins. 
%On the other hand, in the  NaI--NaD plot [panel (a)] the predicted IMF slope from the MIUSCAT models 
%for the most massive bin, steeper than the one predicted by the CvD12 models, is in disagreement with the other indicators, 
%other spectroscopic studies (\citealt{Conroy2012b}), and gravitational lens observations (\citealt{Treu2010, Spiniello2011, Spiniello2012, Barnabe2013}). 
% In fact in panel (a) the most massive bin requires an IMF slope as steep as $x$ = $3.5$ and super-solar metallicity ($Z=0.22$), 
%whereas in panel (b) it requires a less steep IMF slope ($x \sim 3.3$) and super-solar metallicity. 

We emphasize that NaD lies in a part of the spectrum for which the two sets of models use the same stellar library, 
while NaI lies in the redder part where the models make use of two different stellar libraries. 
Therefore the large difference in the NaD behavior cannot be attributed to the different stars used when constructing 
the SSP models and is indeed likely due to a varying [Na/Fe] abundance. 

In conclusion, we argue that the observed trends of NaD and NaI are unlikely to be completely due to a variation in the 
IMF as a function of velocity dispersion but may (in part) be caused by Na-enhancement in massive galaxies.

%{\bf The behavior of the sodium spectral features in massive galaxies has received significant attention recently, 
%as some very massive galaxies show enhanced NaD and NaI (\citealt{Jeong2013}).

\section{IMF slope versus velocity dispersion}
\label{sec:imfsigma}
We now give a more quantitative expression for the variation of the IMF slope with stellar velocity dispersion, following the same approach as in S14.  
In particular, we compare each stacked SDSS spectrum with grids of interpolated and extended SSPs. 
The models cover a large range of ages (log(age) = [0.8 -- 1.15] Gyr, with a step of 0.01 Gyr), [$\alpha$/Fe] 
(between $-0.2$  and $+0.4$ dex, with a step of 0.05 dex), total metallicity ([M/H] between -0.4 and +0.22) 
and IMF slope (x = [1.8 -- 3.5], with a step of 0.1). 
%For each galaxy spectrum we compute $\chi^{2}$ value for both the CvD12 and the MIUSCAT extended models.
The following indices: H$\beta$, Mgb, Fe5270, Fe5335, bTiO, aTiO, TiO1, TiO2, CaH1, CaT1, CaT2, CaT3 allow us 
to constrain the IMF slope and concurrently break the degeneracies between age, abundance ratio and total metallicity. 
For each velocity-dispersion bin and for each SSP model, we compute the $\chi^{2}$ and then 
obtain the best-fitting IMF slope and its uncertainty, marginalizing over age, metallicity and [$\alpha$/Fe]c and 
assuming flat priors on all parameters.
%to obtain a best-fitting IMF slope and its uncertainty (1-$\sigma$ error on the cumulative probability distribution) for each velocity-dispersion bin, 

Unfortunately we cannot investigate the effect of changing the effective temperature of the isochrones, as we did in S14, because MIUSCAT models 
with $\Delta T_{\rm eff} \neq 0$ are not publicly available. However, we perform a test for CvD12 to provide a comparison 
to the non-extended CvD-based fits performed in S14. 
We repeat the $\chi^{2}$ analysis with the same set of indicators for CvD12 SSPs with varying age,  [$\alpha$/Fe], 
[M/H], IMF and $\Delta T_{\rm eff}$ (between $-200$ K and $+200$ K, with a step of 50 K). We find that age, [$\alpha$/Fe] and 
IMF constraints are almost independent  of $\Delta T_{\rm eff}$ (i.e. results obtained with models with 
$\Delta T_{\rm eff} = 0 $ are consistent with those obtained with $\Delta T_{\rm eff}$ as free parameter), 
whereas inferred total metallicities are systematically lower when the $\Delta T_{\rm eff}$ is allowed to change.
We note that compared to S14, where a trend of changing the temperature scale of the giant stars with galaxy mass was 
inferred using models with solar metallicity  (more massive galaxies seemed to have a colder population), here all SDSS bins are consistent
with a $\Delta T_{\rm eff} =-50 \pm 30$.

\begin{figure}  
\center
\includegraphics[height=9cm]{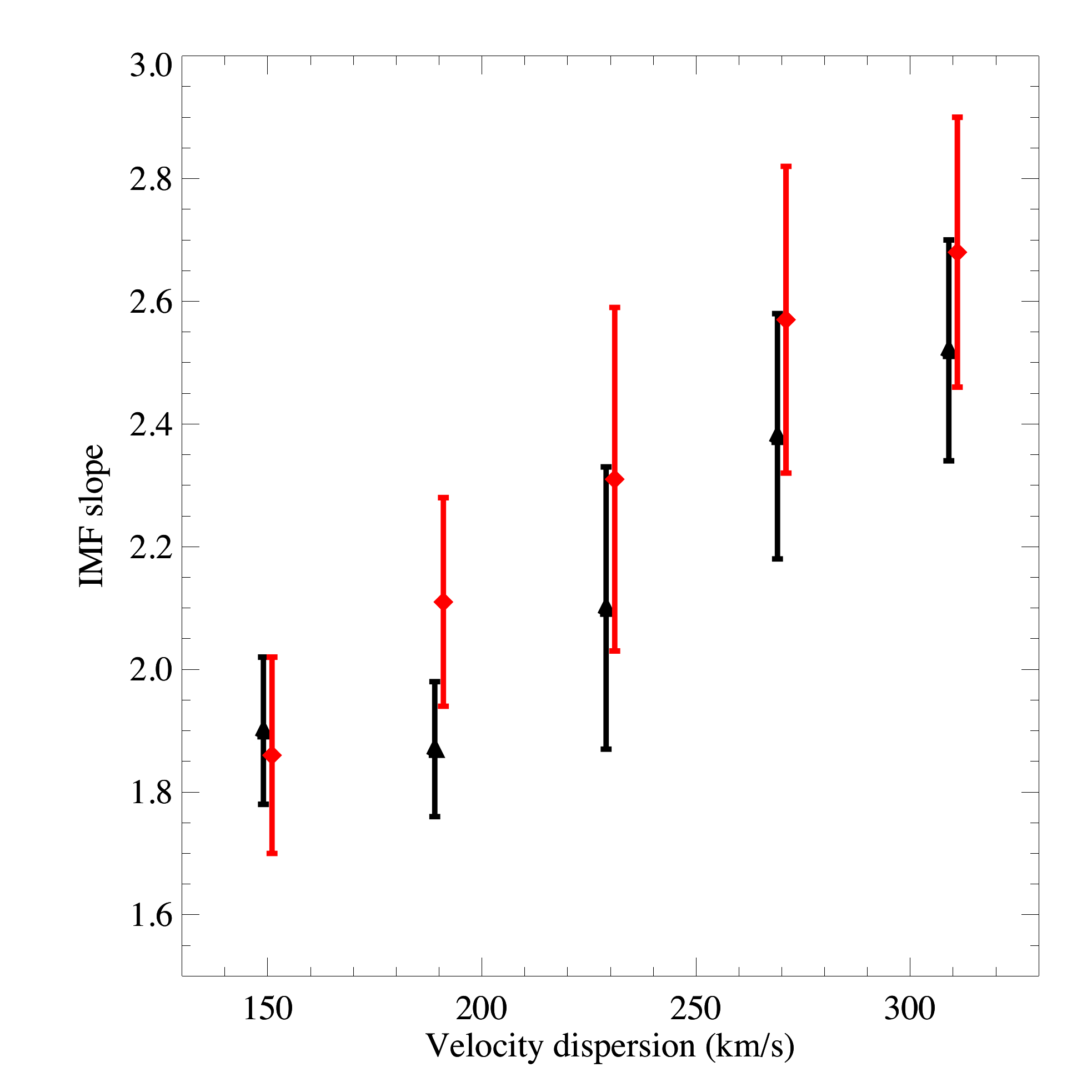}
\caption{Variation of the IMF slope as a function of stellar velocity dispersion predicted from the CvD12 SSP models (black) and the MIUSCAT SSP models (red). 
Points are SDSS ETGs stacked by velocity dispersion. A fair agreement (within 1$\sigma$ error) is found between the IMF slopes predicted from the two sets of models and also with previous work (e.g. S14, \citealt{LaBarbera2013}.}
\label{fig:ssp_gradients}
\end{figure}

We plot the IMF--$\sigma$ relation in Figure \ref{fig:ssp_gradients} for both models (CvD12 in black, MIUSCAT in red).
A remarkable agreement is found between the IMF slope inferred on each SDSS spectrum from the two different set of models, although 
MIUSCAT predicts on average slightly-steeper  (by about $+0.2$) IMF slopes for the more massive bins.
The result of our analysis confirms the findings of S14 and shows that the non-universality of the low-mass end of the IMF
 is a robust and model independent result.

\section{Conclusions}
\label{sec:ssp_conclusion}

In this paper we have compared the two state-of-the-art of SSP models by \citet{Conroy2012} and \citet{Vazdekis2012}, 
specifically constructed for the purpose of studying the stellar  population of old, metal-rich systems, with SDSS ETGs galaxies 
with increasing stellar velocity dispersions from $150$ to $310\,\kms$. We show that both models predict a non-universal low-mass 
end of the IMF slope that steepens with  increasing galaxy mass. 
%qualitatively similar steepening of the IMF slope with increasing galaxy mass. 
%This result demonstrates the robustness of the IMF--$\sigma_{\star}$ variation found in \citet{Spiniello2014}. 

%The parameter-space investigated by the public version of each model could be limiting when trying to
%break the degeneracies between IMF variation and variation of other stellar population parameters, especially in the case of massive ETGs.}
To overcome the limits of each model, we have extended their parameter spaces 
by calculating two independent response functions (one based on metallicity from MIUSCAT 
and another based on [$\alpha$/Fe] from CvD12) to better 
study massive elliptical galaxies, which are overabundant in $\alpha$-elements relative 
to the Sun and metal-rich (e.g., \citealt{Peterson1976,Peletier1989,Worthey1992b,Trager2000, Arrigoni2010}). 

Although this model extension is not fully consistent because of the different ingredients and 
assumptions of the SSP models, we show that it is very useful to demonstrate the
robustness of the results on the IMF variation obtained with CvD12 or MIUSCAT models separately.
Moreover, using these two response functions, we find remarkable agreement between these two SSP models outside their original 
parameter space when the Na lines are excluded, something that a priori might not have been expected.
 
Our main conclusions are the following:
\begin{itemize}
\item Independently of the chosen stellar population model, the non-universality of the low-mass end of the IMF 
is a robust result. A clear trend of the IMF slope with galaxy velocity dispersion is found, under the assumption of  a universal \Mcutoff. 
This result is consistent with other published works (\citealt{Treu2010, Spiniello2012, Cappellari2012, Tortora2013, LaBarbera2013}, S14)

 % showing TiO$_2$, bTiO and CaT, and predictions from the two different models are consistent. 
%, except for the plot NaI – NaD which cannot be used to constrain the IMF.

\item The possibility of exploring super-solar  [$\alpha $/Fe] and super-solar metallicity at the same time is important to break degeneracies
in the stellar population parameters and constrain the low-mass end of the IMF slope, especially for massive ETGs.  
We have enabled this by defining two independent response functions that allow us to extrapolate these SSP models 
beyond their original parameter spaces, finding remarkable agreement using either of the two functions with their respective SSP model.

%\item The data do not require strongly super-solar [$\alpha $/Fe] models, and  
%we conclude that more massive systems are in general slightly more metal-rich 
%(e.g.\ \citealt{Trager2000, Arrigoni2010}). 
\item We find a good agreement with the IMF--$\sigmastar$ relation obtained in S14 using the extended version 
of the CvD12 models that allows us to vary age,  [$\alpha$/Fe], IMF, $\Delta T_{\rm eff}$ and [M/H]. 
However the previously-reported trend of a growing deviation of the temperature of the red giant branch from that predicted by the isochrones
 with galaxy mass disappears when metallicity is taken as free parameter, 
 and all $\sigma$--bins are consistent with a $\Delta T_{\rm eff} =-50 \pm 30$.
 
 \item The indices bTiO, TiO$2$, CaH1 and CaT are robust tracers of the IMF slope. 
Their index strengths give the same predictions for the IMF slope from  the extended version of the two models, 
 and there is a minimal dependence on age (at least for old ages), metallicity and [$\alpha$/Fe]. 
TiO$2$ plotted as a function of CaH1  allows us to break the [$\alpha$/Fe] -- IMF degeneracy since the variation of IMF 
slope is orthogonal to the [$\alpha $/Fe] enrichment in this particular plot. 
However, we note that a zero-point offset is present between the EWs of almost all indicators predicted by the two sets of SSP models. 
Several possible causes for these offsets are investigated in the Appendix but further investigation is required. 
We nevertheless note that the zero-point shifts do not change our main conclusion that the slope of the low-mass end of the IMF is not universal.

 \item The situation is more complicated for  the indices NaD and NaI. 
Specifically, the two different models give different predictions for the IMF--$\sigmastar$ relation 
when these indices are considered. 
Observed  NaD--NaI values only match CvD12 SSP models with solar metallicity and abundances  for ETGs with $\sigmastar  < 250\,\kms$, 
while MIUSCAT models with varying metallicity and solar-abundances  match the data in all mass bins (Fig.~\ref{fig:ssp_data2}) 
but predict an extremely strong variation of the IMF slope, with a maximum of $x$ = $3.5$ for the most massive mass bin. 
This result contradicts inferences from the other indices and other published results 
(\citealt{Spiniello2011, Spiniello2012, Spiniello2014, Cappellari2012, Conroy2012b, Barnabe2013}). 
We therefore conclude that the use of Na indices to constrain 
the IMF slope should be carefully examined and treated with caution.  
Sodium indices should never  be used by themselves to constrain the IMF, 
particularly if one limits oneself to solar-scaled models. 
NaD is especially affected because it is strongly dependent on [Na/Fe] abundance (at least in CvD12). 
The strong difference in the NaI index remains unexplained, even when allowing for non solar Na-abundances. 

%For CaT, CvD12 models with solar [$\alpha $/Fe] provide the best match with the data (in all mass bins). 
%This suggests that the stellar coverage of the IRTF library is perhaps more representative 
%of low-mass stars than the CaT library used by MIUSCAT.
%\item CvD12 models with [Na/Fe] = $0.4$ -- $0.5$ dex better reproduce higher-mass ETGs 
%that are indeed Na-enhanced systems  (i.e.,\citealt{Jeong2013}) with a Salpeter, or slightly steeper IMF slope. 
\end{itemize}

%We investigated the reasons for this disagreement and found that 
%NaI is $\sim 4$ times more sensitive to variations in IMF slope than to variations in sodium abundance, 
%while NaD is $\sim3$ times more sensitive to [Na/Fe] than to IMF slope. 
%Although the models also differ in the NaI--TiO2 plot, if one assumes that giant ETGs are Na-enriched (\citealt{Jeong2013}), 
%then the prediction for the IMF slope variation are comparable and milder than in the NaD--NaI plot. 
%If we assume that iant ETGs are Na-enriched (\citealt{Jeong2013}),  then the NaI--TiO2 plot agrees with conclusions of previously published works 
%Moreover, both models (if we only consider solar Na abundance) predict an extremely strong variation of the IMF slope, 
%with a maximum of $x$ = $3.8$ for the most massive mass bin. 
%This result goes against inferences from all the indices and other published results 
%(\citealt{Spiniello2011, Spiniello2012, Cappellari2012, Conroy2012b, Spiniello2013, Barnabe2013}). 
%since they are affected by still-not-well understood problems. 
%The NaI--TiO2 plot agrees with conclusions of previously published works 
%(\citealt{Cappellari2012, Conroy2012, Spiniello2012, Smith2012, Tortora2013}). 

We note that individual elemental abundance variations should be 
further explored to isolate and test a possible variation of the low-mass end of the IMF slope with galaxy mass.
A full-spectrum fitting approach should be the final goal to investigate possible IMF variations with stellar velocity 
dispersion or other galaxy parameter (such as mass or density)  and to disentangle IMF from age, 
metallicity and elemental abundances (\citealt{Conroy2013b}).  
However the approach taken in this paper (and in S14), the first to attempt to compare the two codes fairly, 
focuses on using specific indices such 
that one better understands how different parts of the spectrum react to changes in age, 
metallicity, abundance ratios, effective temperature and the IMF slope. 
In addition, it avoids potential issues with flux calibration.

In conclusion, all IMF-sensitive indicators in both models give support to the idea of a non-universality of 
the low-mass end of the IMF slope, which increases with increasing galaxy mass.
Using either CvD12 or MIUSCAT SSP models, a bottom-light IMF such as the Milky Way IMF is inappropriate 
for the most massive ETGs, as also shown in \citet{Spiniello2012, Spiniello2014, Barnabe2013}. 
A similar conclusion has been reached in a completely independent way 
using very different approaches such as dynamics or gravitational lensing analyses (\citealt{Treu2010, Cappellari2012}). 

\section*{Acknowledgments}
We thank Charlie Conroy for providing alternate versions of his models and assistance interpreting them. 
C.S. thanks Matteo Barnab\`e, Claudia Maraston, Daniel Thomas, Amina Helmi,  Gerjon Mensinga and Gergo Popping 
for thoughtful comments that have helped to improve the quality of the manuscript.
We thank Michele Cappellari, Sukyuong Ken Yi and Hyunjin Jeong for 
interesting discussions and suggestions.
%The authors thank Charlie Conroy for assistance and help with using and interpreting his models.
Funding for SDSS-III has been provided by the Alfred P. Sloan Foundation, the Participating Institutions, 
the National Science Foundation, and the U.S. Department of Energy Office of Science. 
The SDSS-III web site is http://www.sdss3.org/.
SDSS-III is managed by the Astrophysical Research Consortium for the Participating Institutions of the SDSS-III 
Collaboration including the University of Arizona, the Brazilian Participation Group, 
Brookhaven National Laboratory, University of Cambridge, Carnegie Mellon University, University of Florida, the French Participation Group, the German Participation Group, Harvard University, the Instituto de Astrofisica de Canarias, the Michigan State/Notre Dame/JINA Participation Group, Johns Hopkins University, Lawrence Berkeley National Laboratory, Max Planck Institute for Astrophysics, Max Planck Institute for Extraterrestrial Physics, New Mexico State University, New York University, Ohio State University, Pennsylvania State University, University of Portsmouth, Princeton University, the Spanish Participation Group, University of Tokyo, University of Utah, Vanderbilt University, University of Virginia, University of Washington, and Yale University.

%%%%%%%%%%%%%%%%%%%%%%%%%%%%%%%%%%%%%%%%%%%%%%%%%%%%%%%%
% INPUT BIBLIOGRAPHY
%%%%%%%%%%%%%%%%%%%%%%%%%%%%%%%%%%%%%%%%%%%%%%%%%%%%%%%%
\bibliographystyle{mn2e_fix}
\bibliography{full_phd_Kiara.bib}

\appendix
\section[Comparing model predictions]{Comparing model predictions}
\label{sec:ssp_comparison}
In this Appendix we show how crucial the underlying model assumptions and ingredients like stellar libraries or isochrones 
(especially at high metallicity and for non-solar abundance ratios)  are in giving quantitative inferences about the low-mass end of the IMF slope. 
 In fact, although the qualitative trends of the IMF-sensitive features in the two models are similar
(both predict an increase (decrease for CaT) in the index strengths 
from a MW-like to a bottom-heavy IMF), the variation is generally milder for CvD12 and can be very different
for certain indices. 

The CvD12 and the MIUSCAT models use two different libraries in the optical red and NIR regions 
(CvD12 uses IRTF; MIUSCAT uses CaT and Indo-US), while 
in the blue region ($3500$--$7400$~\AA) they use the same empirical spectral library (MILES). 
%although in slightly different ways. 
The new set of blue IMF-sensitive indicators defined and used in S14 
are therefore essential in this context to eliminate differences arising from the use of different libraries. 
CvD12 and MIUSCAT predict slightly different variations of indices with IMF slope 
even in the wavelength region where they make use of the same empirical spectral library. 
We investigate the reason of this disagreement and find that one of the largest differences 
is the use of different isochrones. When using  CvD12 models made with the same isochrones used in MIUSCAT models,  
we find better agreement between the models. 
However, even when the two sets of models use the same library and the same isochrones, 
we still find small differences in the predictions of the IMF slopes for some indices. 
We attribute to the different methods that the CvD12 and the MIUSCAT models
use to attach stars to the isochrones and also to a possible mismatch in the assumed effective temperature of cold stars.  
Both models test different assumptions for the IMF shape and are created for the specific purpose 
of examining the stellar content of massive ETGs. %, despite still both being calibrated on the solar neighbourhood. 
They are both mainly based on empirical libraries that generally provide good fits to 
line-strengths and full spectra of populations of Solar neighborhood stars. %(e.g.\ \citealt{Carter1986}).
However, empirical stellar libraries are often not able to reproduce consistently 
the spectral features of systems which have undergone a star formation history 
(SFH) different than the solar-neighborhood. This is for instance the case of ellipticals, 
which have been shown to be overabundant in $\alpha$-elements with respect to the Sun 
(e.g., \citealt{Peterson1976,Peletier1989,Worthey1992b}).
This happens because, by construction, the abundance pattern of models based on empirical libraries 
is set by the stars in the library, which are mainly observed in the 
solar neighborhood. % (\citealt{McWilliam1997}).
On the other hand, a clear advantage of using real stars is that they do not rely on our knowledge 
of the physics of stellar atmospheres and databases of atomic and molecular transitions.
%For instance, the empirical spectra contain the effect of stellar winds that affect the photospheric 
%lines of massive stars and are too complicated to model in theoretical libraries (\citealt{Kudritzki1987}).
%In particular, model are not able to well reproduce $U-B$ and $B-V$ colors, expecially for 
%cool stars ($T_{\rm eff} < 4000$K). 

\subsection{Isochrones}
An important difference between the models is the set of isochrones 
used to calculate stellar parameters and spectra.
A large number of isochrones exist in the literature (see \citealt{Conroy2013} for a review), 
spanning a wide range of ages and chemical compositions for stars with masses between 
the hydrogen burning limit ($\sim 0.08\,M_{\odot}$) and $\sim 100\,M_{\odot}$.
Different sets of isochrones are tailored for different mass ranges and different evolutionary phases of stellar evolution. 
Some are more effective in tracking the high-mass stars, 
others focus on the main sequence, red giant branch (RGB), and horizontal branch (HB) 
evolution of low-mass stars, such as the Dartmouth models (\citealt{Dotter2008}). 
Others are particularly effective at describing the very low-mass end of the IMF down to the brown dwarf regime, 
such as the Lyon models (\citealt{Chabrier1997, Baraffe1998}). 
%The most challenging evolutionary phase to cover is without doubt the post-AGB phase. 
Since no single set of available isochrones covers a full range of ages, 
metallicity and evolutionary phases, most stellar population synthesis 
models use a combination of different isochrones. 
Combining various sets of isochrones is not trivial because different stellar interior models
can make different physical assumptions (convection, rotation, etc.), 
and consequently the age at which stars evolve away from the main sequence can vary between models.\\
\indent Commonly-used isochrones for the bulk of the age and metallicity 
range of elliptical galaxies include the Padova isochrones (\citealt{Bertelli1994, Girardi2000, Marigo2008}) 
 and the BaSTI models (\citealt{Pietrinferni2004, Cordier2007}). These are often
supplemented with the Geneva models (\citealt{Schaller1992, Meynet2000}) at younger ages. 
Little attention has been paid in the past to the lowest-mass portion of the isochrones, 
since low-mass stars contribute only few percent of the light of an old stellar population 
in the optical (e.g.\ \citealt{Worthey1994, Renzini2006, Conroy2013}). 
However, if one wants to study the low-mass end slope of the IMF, this 1--5\% 
contribution is crucial. It is for this reason that 
CvD12 decided to use the Lyon models for stars with masses below 
$0.2\,M_{\odot}$. %while MIUSCAT models use Padova isochrones (\citealt{Girardi2000}) everywhere. 
The Lyon models use the surface boundary condition of the base of the atmosphere 
(rather than at $T$ = $T_{\rm eff}$) which is better for stellar interior codes that 
are not ideally suited to compute the physical conditions for high-density, 
low-temperature environments down to $T$ = $T_{\rm eff}$ (CvD12).
Using only the solar-scaled theoretical Padova isochrones of \citet{Girardi2000}, the MIUSCAT 
library cuts off at $0.15\,M_{\odot}$, since the Padova isochrones 
do not extend to lower masses. 

The CvD12 models allow a more accurate treatment of the low-mass 
stars down to the hydrogen burning limit (\Mcutoff $= 0.08\,M_{\odot}$), 
using several separate evolutionary calculations (a combined set of different isochrones, empirical and theoretical libraries)
and adding the SDSS dM stellar templates of \citet{Bochanski2007} to the MILES library. 

In Figure~\ref{fig:ssp_data1} an offset between the EWs of TiO lines predicted from MIUSCAT and CvD12 models with same parameters is found. 
To investigate the origin of this offset and to test the importance of isochrones in constraining the IMF slope from TiO indices, 
Prof. Conroy kindly built and provided us with new CvD12 models using Padova isochrones for all stellar evolutionary phases. 

\begin{figure*}
\centering
\includegraphics[height=7 cm]{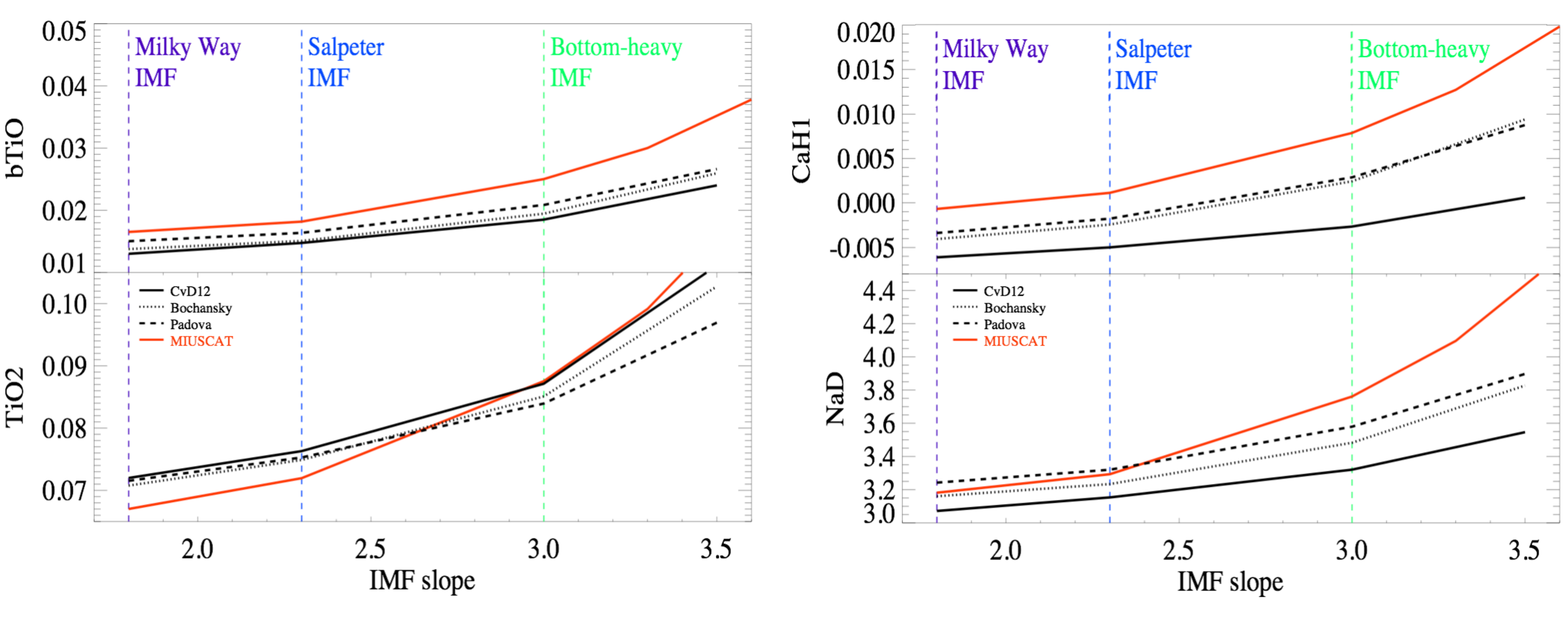}
\caption{ Variation of bTiO, TiO$2$, CaH1 and NaD indices with IMF slope. 
Black solid lines are a CvD12 SSP model of 12.5 Gyr built using different set of isochrones 
for  the different stellar evolutionary phases, Black dotted lines are the same models 
with the addition of the SDSS dM stellar templates of  \citet{Bochanski2007} 
to augment the numbers of M dwarfs. Black dashed lines are CvD12 SSP models built using the Padova isochrones 
(see caption and text for more details). 
Red lines are MIUSCAT SSP models with the same age. 
Vertical colored lines show different IMFs. }
\label{fig:ssp_variationCvD12}
\end{figure*}

In Figure~\ref{fig:ssp_variationCvD12} we show the response of the EWs of IMF-sensitive features to the variation of the low-mass 
end of the IMF slope for the publicly available versions of MIUSCAT (red solid line) and CvD12 (black solid) SSP models 
with the same age, solar [$\alpha $/Fe] and metallicity, and for the modified CvD12 models with Padova isochrones (dashed black line). 
We also plot a CvD12 model with the same age and metallicity obtained with the addition of the SDSS dM stellar 
templates of  \citet{Bochanski2007} (dotted line).  Because CvD12 and MIUSCAT make use of the same empirical 
stellar library (MILES) in this spectral region the main difference between the two sets of models is the assumed isochrones.
Although the predictions of the two SSP models agree qualitatively, the 
MIUSCAT models suggest a steeper variation of all the indices with IMF slope, which affects
the IMF--$\sigma^{\star}$ relation. For all indices except TiO$2$, the discrepancy between 
MIUSCAT and CvD12 becomes smaller when the two models make use of the same stellar isochrones, as expected. 

However, in the particular case of the TiO$2$ index, the models do not predict the same variation of  the  indices with IMF slope, 
even when they make use of the same empirical stellar library and set of isochrones and actually 
the CvD-Padova models are in even worse agreement with MIUSCAT Padova-based models. 
Consequently, we argue that this difference must arise from a different cause.  
One possibility is the different methods used in the CvD12 and MIUSCAT models to attach stars to the isochrones at low mass. 
The MIUSCAT models obtain stellar fluxes from the theoretical parameters of the isochrones ($T_{\rm eff}, \log{g}, {\rm [Z/H]}$) 
using empirical relations between colors and stellar parameters 
(temperature, gravity and metallicity, respectively). 
Their algorithm,  described in \citet{Vazdekis2003, Vazdekis2010}, finds the closest stars and weights them according 
to the distance to the target point ($T_{\rm eff,0}$, $\log{g_{0}}, [{\rm Fe/H}]_{0}$) in the stellar parameter space. 
The CvD12 SSP models use instead a $M$--$L_\mathrm{bol}$ relation, choose the closest observed stellar spectrum 
from the IRTF library with the appropriate bolometric luminosity and then 
match it to a MILES spectrum at shorter wavelengths.  For instance, this different interpolation scheme and the 
different physical approach could  cause zero-point shift in the assumed effective temperature of the stars used in the SSP models.
We further discuss this point in the next section, although we do not have the possibility to investigate this in detail.

\subsection{Effective Temperature}
\begin{figure*}
\centering
\includegraphics[height=3.6cm]{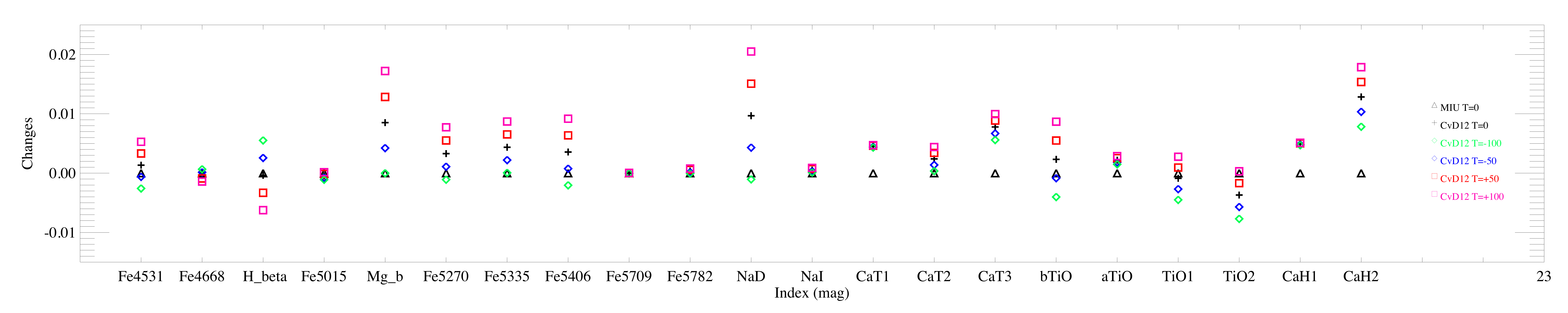}
\caption{Variation of EWs (measured in mag) for CvD12 models with $12.5$ Gyrs, Salpeter IMF, 
solar abundance pattern and varying $T_{\rm eff}$ (different colors, see caption)
with respect to the EW of a MIUSCAT model with same parameters and $T_{\rm eff}=0.$ (black triangle, set to zero value). }
\label{fig:ssp_Teffvariation}
\end{figure*}

We have made an attempt to understand the origin of the zero-point differences of index strengths between the two SSP models 
by changing the temperature of the isochrones as described by Conroy \& van Dokkum (2012a,b), in which the effective 
temperatures of all stars along an isochrone are shifted by an amount within the observational and systematic uncertainties 
in the effective temperature scale of normal giants and dwarfs (roughly $\pm100\,\mathrm{K}$).  
There are not currently public versions of the MIUSCAT models that include such a shift, so we have shifted the effective 
temperatures of a solar abundance, 12.5 Gyr old, Salpeter IMF CvD12 model by up to $\pm100\,\mathrm{K}$ 
and compared these with the same MIUSCAT model\footnote{The effective temperature difference vector 
for $\lambda>4500\,\mathrm{\AA}$ was kindly provided by Dr.~Conroy.}.  We plot the results in Figure~\ref{fig:ssp_Teffvariation}, 
where we have converted indices in \AA\ to magnitudes using the conversion given by Kuntschner (2000, his Eq.\ 2) for easy comparison.  
We find that there is no consistent shift that brings all indices into agreement between the models, and for some indices 
(namely CaT1, CaT3 and the two CaH indices) there is no reasonable temperature shift able to bring 
the two models into agreement.  Without having the ability to change the temperature of stars independently
 along the isochrones, we cannot proceed further with this analysis.  
We merely point out here that these zero-point differences likely arise from the different 
$T_{\mathrm{eff}}$--$M_{\mathrm{bol}}$ relations — i.e., the different methods of attaching 
stellar spectra to isochrones — assumed by the two models and possibly by the different spectral libraries used redwards of 7500 \AA.

\subsection{Index variation comparisons}
\begin{figure*}
\centering
\includegraphics[height=20cm]{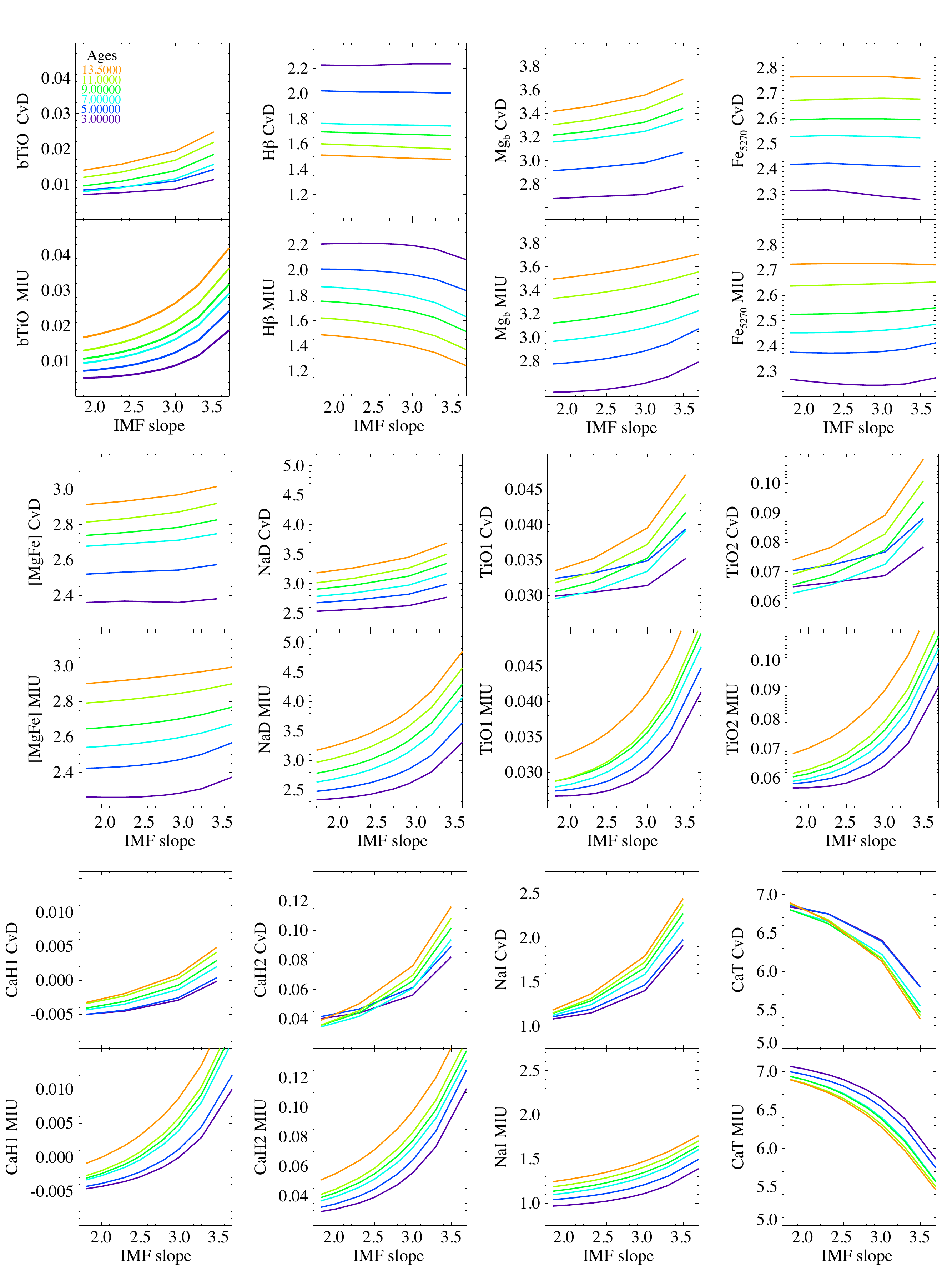}
\caption{Variation of index strengths with IMF slope predicted for the CvD12 models 
(upper panels) and the MIUSCAT models (lower panels),  
convolved to a common resolution of $\sigma$ = $350\,\kms$. 
In each panel, different colors represent SSP with different ages, as indicated 
in the legend on the first panel. TiO and CaH indices are given in magnitudes, 
while all the other indices are given in \AA. Note here that the MW IMF results are given at an IMF slope of $x=1.8$ for convenience.  }
\label{fig:ssp_variationIndex_mc}
\end{figure*}

To compare predictions from the two models, 
we calculate index variations as a function of the IMF slope and the age 
of several optical indices for both the MIUSCAT and CvD12 models. 
In Figure~\ref{fig:ssp_variationIndex_mc}, we show the predicted index variations 
for a range of CvD12 models (upper panels) and MIUSCAT models (lower panels) 
with different ages (lines of different colors). 
%Because of the different approach in dealing with metallicity and [$\alpha$/Fe], 
Here we restrict our comparison to solar abundances and metallicity (using only the public available set of models),  
because here in the Appendix we do not compare the SSP models with massive (metal-rich and $\alpha$-enhanced) ETGs. 
We also restrict the comparison to the unimodal IMF case, the only one explored by CvD12. 

Figure~\ref{fig:ssp_variationIndex_mc} confirms that 
most of the blue classical Lick indices (\citealt{Burstein1984, Worthey1994, Trager1998})  
do not depend (or depend only weakly) on the IMF slope, 
while the Na, TiO and CaH indices increase from MW IMF to a bottom-heavy IMF 
for both models (as already shown in S14). 
The only IMF-sensitive index that grows weaker with increasing IMF slope is  
CaT, as seen in previous studies (e.g.\ \citealt{Cenarro2003, Conroy2012}). 

For the majority of the indices, the gradient of the variation is similar, although generally slightly 
milder for the CvD12 models than for the MIUSCAT models. 
For some indices, however, the two SSP models give quite different predictions for IMF variations.
CaH1 is an extreme case, and in addition bTiO and both the sodium indices 
(NaI and NaD) behave differently in the two sets of models.
We further investigate the behavior of these particular indices in the following sections. 
Predictions for the H$\beta$ index are also different, although this index depends only weakly on the IMF slope. 
In fact, for the MIUSCAT models the H$\beta$ decreases with increasing IMF slope, whereas it remains
nearly constant for CvD12 models. Because this index is mainly (but not entirely: see Worthey,1994) 
contributed by turn-off stars at solar metallicity (\citealt{Buzzoni1994}), its sensitivity to the IMF slope must
be understood as a relative change of the contributing fraction of such hot
stars.  

On the other hand, index variations with IMF slope of the redder TiO features predicted 
from the two SSP models are similar, and variations of CaH2 and CaT are also similar. 
%The only index that shows a steeper variation with IMF slope in the CvD12 model is NaI ($\lambda 8190$ \AA );
However, for these indicators, the trends of the variation of EWs with age  reverse for the youngest ages. 
A possible explanation for this could be the presence of a more extended AGB contribution in the CvD12 
models with respect to the MIUSCAT models.

Moreover, for some indices, there is also a shift in the zero-point: 
for the MW IMF, the NaD indices of the CvD12 models with {\sl solar abundance} are systematically lower than the MIUSCAT 
models while the TiO$1$ and TiO$2$ indices are systematically higher.
Small differences in the metallicity-sensitive features could be due to the different 
ways the two sets of models deal with metallicity and/or [$\alpha $/Fe]. 
This will be addressed in the following section.

To further quantify the differences between the variation with IMF 
predicted on a single index from the two sets of SSP models we compute the following 
quantities, 

\begin{eqnarray}
\,\,\,\,\,\,\,\,\,\,\,\,\,\,\,\,\,\,\,\,\,\,\,\,\,\,\,\,\,\,\,\,\Delta {\rm IMF} \hphantom{Sal} &= &I_{i,(x=3.5)} -  I_{i,({\rm MW})} \\
\,\,\,\,\,\,\,\,\,\,\,\,\,\,\,\,\,\,\,\,\,\,\,\,\,\,\,\,\,\,\,\,\Delta {\rm IMF_{Salp}} &= &I_{i,(x=3.5)} -  I_{i,({\rm Salp})} \\
\,\,\,\,\,\,\,\,\,\,\,\,\,\,\,\,\,\,\,\,\,\,\,\,\,\,\,\,\,\,\,\,\Delta {\rm IMF_{{\rm MW}} }&= &I_{i,({\rm Salp})} \hphantom{S}-  I_{i,({\rm MW})} 
\end{eqnarray}

for the index $i$, in an old model with an age of $\sim 13.5$ Gyr for CvD12 and 
$\sim 14.1$ Gyr for MIUSCAT. 
The first equation measures the variation of the index $i$ with IMF slope 
from the MW IMF to an extremely bottom-heavy IMF (slope of $x = 3.5$), 
while the second and the third measure the variation of the index $i$ with IMF slope 
from a Salpeter to $x = 3.5$ and from a  MW IMF to a Salpeter IMF respectively.
%Thus, the values $\Delta IMF$ indicate a measure of the sensitivity of a given index to IMF steepening: 
%indices with higher values of $\Delta IMF$ are more  suitable to constrain the IMF slope. 

In Table~\ref{tab:comparison_delta} we report the values of the three $\Delta {\rm IMF}$s for the selected indices, separated
by the units in which they are computed. 
We also report between parentheses the fractional changes (for indices in \AA, these are calculated by normalizing each quantity 
with respect to the lower-slope value; for magnitude indices, we use the approximation that for small changes, i.e. $<25\%$, 
magnitudes differences are nearly the same as percentage changes).

\begin{table*}
\center
\caption{$\Delta {\rm IMF}$ for the selected indices predicted from the two SSP models. Indices are separated by 
the units in which they are computed. We report as reference the value of the indices for a model with a Salpeter IMF. This values
are useful to highlight zero-point differences in the two sets of models. 
We report the percentage change compared to the EW for the lower slope IMF in each case between parentheses. }
\begin{small}
\begin{tabular}{lrrrrrrrr} 
\hline
\hline
\smallskip
\textbf{Index} & \textbf{EW$_{Salp}$} & \textbf{EW$_{Salp}$} &\textbf{$\Delta {\rm IMF}$} & \textbf{$\Delta {\rm IMF}$} & \textbf{$\Delta {\rm IMF_{Salp}}$} &\textbf{$\Delta {\rm IMF_{Salp}}$} & \textbf{$\Delta {\rm IMF_{{\rm MW}}}$} & \textbf{$\Delta {\rm IMF_{{\rm MW}}}$} \\
\textbf{(mag)} & \textbf{CvD12} & \textbf{MIUSCAT} & \textbf{CvD12} & \textbf{MIUSCAT} & \textbf{CvD12}  & \textbf{MIUSCAT}& \textbf{CvD12}& \textbf{MIUSCAT} \\
\hline
 bTiO	    & 	0.016   &   	0.019 &		0.011(1.1\%)	&	0.026(2.6\%)	&		0.009(0.9\%)	&		0.023(2.3\%) 	&		0.002(0.2\%)	&			0.003(0.3\%)   \\
 TiO$1$  	&  0.035   &		0.034    &    	0.014(1.4\%)	&	0.017(1.7\%)	&		0.012(1.2\%)	&		0.015(1.5\%)	&		0.002(0.2\%) 	&	        0.002(0.2\%)  \\ 
 TiO$2$	&	0.078  	&		0.074    &		0.034(3.4\%)	&	0.048(4.8\%)	&		0.030(3.0\%) 	&		0.042(4.2\%)	&		0.004(0.4\%) 	&		    0.005(0.5\%)   \\  
 CaH1	    & $-$0.0019  &	  0.0017    &		0.008(0.8\%)	&	0.020(2.0\%)	&		0.006(0.6\%) 	&		0.018(1.8\%) 	&		0.001(0.1\%)  	&		    0.003(0.3\%)   \\   
 CaH2	   &	 0.050  &   	0.064    &   	0.077(7.7\%)	&	0.095(9.5\%)	&		0.066(6.6\%) 	&		0.082(8.2\%)	&		0.011(1.1\%) 	&		    0.001(0.1\%)   \\    
\hline 
\textbf{Index} & \textbf{EW$_{Salp}$} & \textbf{EW$_{Salp}$} &\textbf{$\Delta{\rm IMF}$} & \textbf{$\Delta {\rm IMF}$} & \textbf{$\Delta {\rm IMF_{Salp}}$} &\textbf{$\Delta {\rm IMF_{Salp}}$} & \textbf{$\Delta {\rm IMF_{{\rm MW}}}$} & \textbf{$\Delta {\rm IMF_{{\rm MW}}}$} \\
\textbf{(\AA) } & \textbf{CvD12} & \textbf{MIUSCAT} & \textbf{CvD12} & \textbf{MIUSCAT} & \textbf{CvD12}  & \textbf{MIUSCAT}& \textbf{CvD12}& \textbf{MIUSCAT} \\
\hline
 H$\beta$ &		1.502 	&    1.461     &	  $-$0.034(2.3\%) 	&	 $-$0.14(9.6\%)	&	  $-$0.024(1.6\%)    &	  $-$0.116(8\%)   &	  $-$0.011(0.7\%)  	&		$-$0.027(1.8\%)   \\    
 Mg$b$     &	    3.460  &     3.535    &  	0.274(8.0\%) 	&	0.153(4.4\%)	&		0.230(6.6\%)	&	  0.113(3.2\%)	&		0.044(1.3\%) 		&		0.040(1.1\%)   \\    
 Fe5270   	 &		2.766  &     2.726   	&	  $-$0.007(0.2\%)   & 0.001(0.05\%)&	 $-$0.009(0.3\%)	&	  $-$0.001(0.05\%) 	&	   	0.002(0.07\%) 		&		0.003(0.1\%)    \\  
 MgFe	 	&		2.932   &  	2.919   &		0.102(3.5\%)	&	0.067(2.3\%)	&		0.083(2.8\%)	&	  0.049(1.7\%)  	&	    0.018(0.6\%) 	&		0.018(0.6\%)   \\     
 NaD   		&	    3.268   &		3.359   &  	0.507(16\%)		&	1.00(31\%) 	&		0.42(13\%)		&		0.816(24\%)		&		0.087(2.8\%) 	&		0.185(5.8\%)   \\  
 NaI			&		1.365   &   	1.313   &		1.26(100\%)		&	0.337(27\%)		&		1.079(79\%)		&		0.267(20\%)   &  	0.182(15\%) 		&		0.070(5.6\%)   \\     
 CaT			&		6.655  &		6.723	&	   $-$1.514(19\%)	&	$-$0.926(13\%)	&	  $-$1.28(17\%)		&	   $-$0.757(11\%)  & 	  $-$0.238(3\%)	&		$-$0.17(2.5\%)   \\
\hline	
\hline

\label{tab:comparison_delta}
\end{tabular}
\end{small}
\end{table*}

A good qualitative agreement is found between the two sets of models for some 
of the indices, although the MIUSCAT models typically predict a larger variation of EWs with IMF slope. 
This result also confirms the existence of the IMF--$\sigmastar$ relation given in \citet{Spiniello2014}. 

By comparing the curves in Figure~\ref{fig:ssp_variationIndex_mc} 
with the entries in Table~\ref{tab:comparison_delta},  
we can further understand differences in the index variations of the two models. 
For instance, in the CvD12 models, H$\beta$ does not depend at all on IMF
 (only $\sim 2\%$ variation from the MW IMF to $x=3.5$), while the MIUSCAT models predict a (mild) anti-correlation 
of the EW of H$\beta$ with a fractional change of the EW of 10\%. 
 
TiO and CaH indices behave similarly in the two models 
%between the MW IMF and the Salpeter IMF
, but MIUSCAT predicts overall 
 steeper variations, especially for very bottom-heavy IMFs. 
The CaT index shows an anti-correlation of the EW with the IMF slope for both models. %, followed by 
%CaH, NaI and TiO lines all of them with $\Delta$(IMF/age)$>1$. This confirm the analysis in S14, where these were the main IMF-sensitive indices used  
%to quantify the non-universality of the low-mass end of the IMF and 
%its relation with stellar velocity dispersion. Moreover, the fact that these indices behave similarly in the two sets of models
% further confirm that our result is robust and model independent.

The values of the variations predicted by CvD12 for the NaI and NaD indices 
are instead very different in all the $\Delta$IMF ranges 
than the values predicted for the same features by MIUSCAT, even though the values for a MW-like IMF are 
similar between the models (i.e. no zero-point shift is present in the case of sodium). 

\section{The [Na/Fe] abundance pattern from CvD12}
In Section~\ref{sec:ssp_data} we conclude that both the CvD12 and MIUSCAT 
extended versions of the models suggest a non-universal IMF whose low-mass end steepens 
with the velocity dispersion for the averaged SDSS ETGs. 
This result clearly shows that the dependence of 
the low-mass end of the IMF ($\le 0.3\,M_{\odot}$) on the stellar velocity dispersion 
(stellar mass) of the system  (e.g.\ \citealt{Treu2010, Auger2010b, 
Napolitano2010, vandokkum2010, Spiniello2011, Spiniello2012, Spiniello2014}) 
is genuine and does not arise from any misunderstanding of the ingredients of the SSP models. 

\begin{figure}  
\includegraphics[height=7cm]{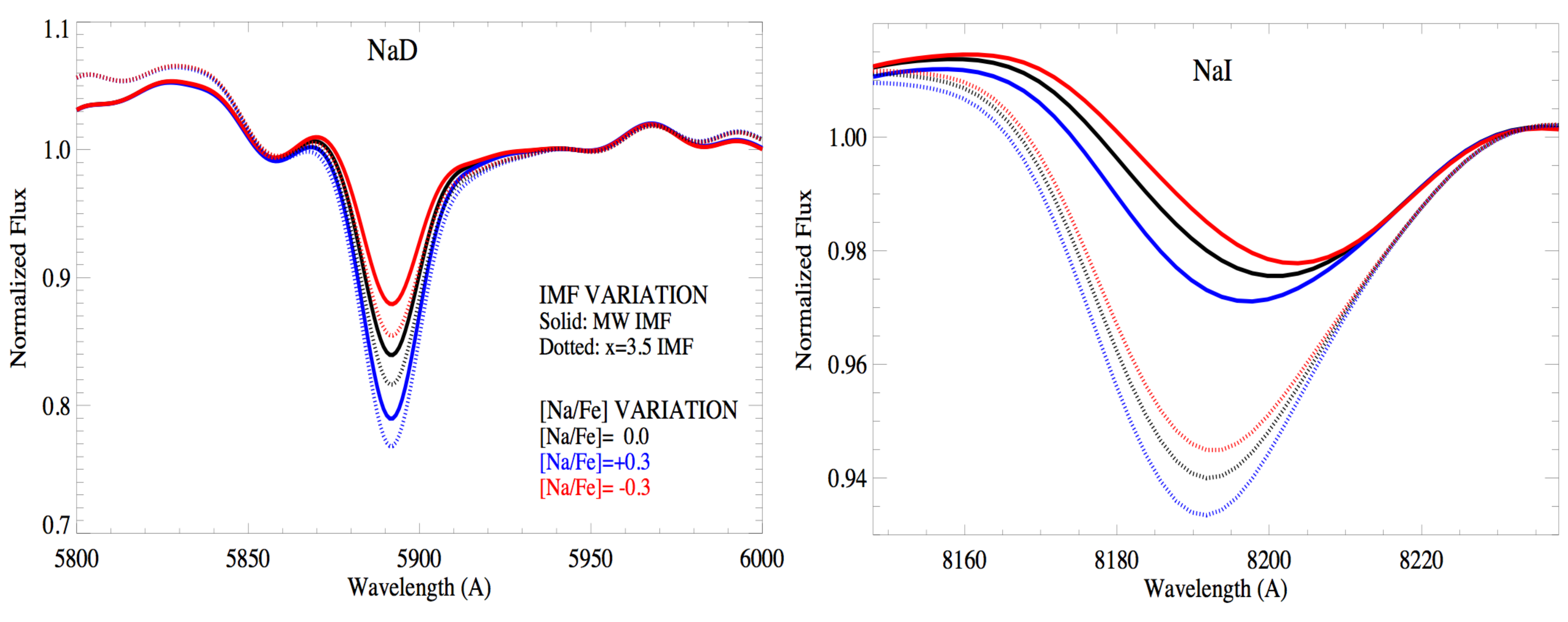}
\caption{Spectra of a CvD12 model with an age of $13.5$ Gyr in
  the regions of the NaD (left) and NaI (right) features. Different colors are models with different [Na/Fe] abundances (between $-0.3$ and $+0.3$), 
solid lines are models with  MW IMF, while dotted lines represent models with an extremely bottom-heavy IMF ($x=3.5$). NaD is more sensitive to abundance variations than to IMF, whereas  NaI is much more sensitive to variations in the IMF slope. Because the MIUSCAT models only have the solar abundance pattern, we cannot repeat this test for those models.}
\label{fig:ssp_sodiumlines}
\end{figure}

However, we also demonstrated that  the two different models give different predictions for the IMF--$\sigmastar$ relation 
when sodium indices are considered.
Observed  NaD--NaI values only match CvD12 SSP models with solar metallicity and abundances 
for ETGs with $\sigmastar  < 250\,\kms$, while MIUSCAT models with varying metallicity and solar-abundances match 
the data in all mass bins (Fig.~\ref{fig:ssp_data2}a) but
they predict an extremely strong variation of the IMF slope, with a maximum of $x$ = $3.5$ for the most massive mass bin 
(and moreover they fail to reproduce the NaD--CaH1 EWs at all sigmas.) 
 Here we focus on this issue. We use the CvD12 models, the only current SSP allowing for a non-solar sodium abundance 
pattern, to decouple IMF variations from abundance variations for each of the two sodium lines. 
%The flexibility of CvD12 in varying [$\alpha$/Fe] 
%as well as the abundance pattern of 11 different elements is 
%particularly important in the case of sodium lines, 
%as we have shown in Section~\ref{sec:sodium}. 
Using the CvD12 models with different [Na/Fe] abundances, we study 
the behavior of the NaD and the NaI indices when varying the IMF slopes at fixed 
sodium abundance and when varying [Na/Fe] at fixed IMF slope.
In Figure~\ref{fig:ssp_sodiumlines} we show the CvD12 models with an age of $13.5$ Gyr, zooming in on
the regions of the NaD (left panel) and NaI (right panel) features. 
In both panels different colors show models with different [Na/Fe], from $-0.3$ to $+0.3$, while 
different line-styles represent models with different IMF slopes, from the MW IMF to $3.5$. 
The figure clearly demonstrates that NaD absorption is more sensitive to [Na/Fe] than 
the redder NaI feature, which varies much more with the IMF slope. 

%\resizebox{0.8\hsize}{!}{$
%$}
This sensitivity can be quantitatively expressed as
\begin{equation}
\left(\dfrac{\Delta{\rm IMF}}{\Delta{\rm [Na/Fe]}}\right) _{i} = \dfrac{I_{i,(x=3.5){\rm [Na/Fe]}=0} -  I_{i,({\rm MW}){\rm [Na/Fe]}=0}}
{\langle I_{i, ({\rm [Na/Fe]}=+0.3)}  -  I_{i, ({\rm [Na/Fe]}=-0.3)}  \rangle_{x=[{\rm MW}-3.5]} } 
\label{eq:sod}
\end{equation} 
for both indices (see Table~\ref{tab:sodium_delta}). In this equation, the numerator measures the variation of an index $i$ 
with IMF slope  from a Milky Way-like IMF to an extremely bottom-heavy IMF (slope of $x=3.5$), 
while the denominator is the average of the index variation with sodium abundances 
in the range ${\rm [Na/Fe]}$=[$-0.3$, $+0.3$] for the two IMFs. 

\begin{table}
\center
\caption{Ratio between the fractional changes in indices with IMF slope and with 
sodium abundances at fixed age, predicted by the CvD12 SSP models. }
\begin{tabular}{lc} 
\hline
\textbf{Index} & \textbf{$\Delta{\rm IMF}/\Delta{\rm [Na/Fe]}$} \\ %&
\hline
NaD & 0.29 \\
NaI & 4.21 \\
\hline
\label{tab:sodium_delta}
\end{tabular}
\end{table}

The larger the value of $\Delta{\rm IMF}/\Delta{\rm [Na/Fe]}$ is, the larger 
the sensitivity to IMF slope is compared with the sensitivity to Na-abundance. 
NaI is $\sim 4$ times more sensitive to variations in IMF slope than to variations in sodium abundances, 
whereas NaD is $\sim 3$ times more sensitive to [Na/Fe] than to IMF slope.

Thus a non-solar [Na/Fe] abundance in massive galaxies could explain the fact that in 
panel (a) of Figure~\ref{fig:ssp_data2} CvD12 models with solar abundances only match the low-mass systems.  
As highlighted by \citet{Conroy2012,Conroy2012b} and by S12, more-massive ETGs require higher [Na/Fe] {\sl and} steep 
(Salpeter or slightly steeper) IMF slopes. % (Figure~\ref{fig:sodium_abu}).  
Probably the IMF slopes inferred from MIUSCAT models from panel (a) appear to be steeper because the change in the NaD EWs is 
attributed totally to IMF variations, as the models have solar [Na/Fe] abundance.
NaD is especially affected in this context because it is strongly dependent on [Na/Fe] abundance ($\sim 4$ times more than the redder NaI).
The strong disagreement between the models regarding the NaI index, which is less  affected in this context, 
remains  unexplained. 

%The models also differ in the NaI--TiO2 plot, panel (b), but if one assumes that giant ETGs are Na-enriched (\citealt{Jeong2013}), 
%then the prediction for the IMF slope variation are comparable and milder than in the NaD--NaI plot. 
%This is explicable by the fact that the NaD absorption is more sensitive to [Na/Fe] than 
%the redder NaI feature, which varies much more with the IMF slope. 
%More details on the non-solar [Na/Fe] abundance for more massive galaxies are showed in the Appendix. 

Hence, we conclude that the use of Na indices in constraining  the IMF slope should  be more carefully examined and considered 
with caution. In particular, they should never be used by themselves to constrain the IMF, if one limits oneself to solar-scaled models.

\label{lastpage}

\end{document}